\documentclass[english,10pt]{article}

\usepackage[T1]{fontenc}
\usepackage[hyperfootnotes=false]{hyperref}
\usepackage{ae,aecompl}
\usepackage[english]{babel}
\usepackage[reqno,fleqn]{amsmath}
\usepackage{eurosym}
\usepackage{amssymb}
\usepackage{amsmath}
\usepackage{amsfonts}
\usepackage{amsthm}
\usepackage{array}
\newcommand{\sym}[1]{\rlap{$#1$}} 

\usepackage{booktabs}
\usepackage{caption}
\usepackage{subcaption}
\usepackage{chngpage}
\usepackage{comment}
\usepackage{dsfont}
\usepackage{eurosym}
\usepackage{graphicx}
\usepackage{hyperref}
\usepackage{float}
\usepackage{lscape}
\usepackage{mathrsfs}
\usepackage{multirow} 
\usepackage{natbib}
\usepackage{pdflscape}
\usepackage{setspace}
\usepackage{tabularx}
\usepackage{tikz}
\usepackage{url}
\usepackage{verbatim}
\usepackage{afterpage}
\usepackage[toc,page,header]{appendix}
\usepackage{hyphenat}
\graphicspath{{"figures/"}}
\usepackage[flushleft]{threeparttable}
\usepackage{longtable}
\usepackage{dcolumn}
\usepackage{lscape}

\usepackage[nobottomtitles*]{titlesec}

\usepackage{scrwfile}
\TOCclone[Appendices for online publication]{toc}{atoc}

\AfterTOCHead[toc]{%
}
\AfterTOCHead[atoc]{%
 \edef\maintocdepth{\the\value{tocdepth}}%
 \value{tocdepth}=-10000\relax%
}

\usepackage[left=1 in,right=1 in,top=1 in,bottom=1 in]{geometry}
\newcolumntype{L}[1]{>{\raggedright\arraybackslash}p{#1}}
\newcolumntype{C}[1]{>{\centering\arraybackslash}p{#1}}
\newcolumntype{R}[1]{>{\raggedleft\arraybackslash}p{#1}}

\makeatletter
\def\@biblabel#1{\hspace*{-\labelsep}}
\makeatother
 \def\independenT#1#2{\mathrel{\setbox0\hbox{$#1#2$}%
 \copy0\kern-\wd0\mkern4mu\box0}}

\title{The Psychological Gains from COVID-19 Vaccination: \\ Who Benefits the Most?
\thanks{\scriptsize {Manuel Bagues, University of Warwick, CEPR, IZA and J-Pal; email: manuel.bagues@warwick.ac.uk; Velichka Dimitrova, University of Warwick; email: velichka.dimitrova.1@warwick.ac.uk. We thank Andrew Oswald, Thiemo Fetzer, Negar Ziaeian Ghasemzadeh, Ines Berniell, Ceren Bengu Cibik, Natalia Zinovyeva, Sonia Bhalotra, Hiba Sameen, Andreas Stegmann, Daniel Sgroi and participants in seminars at the University of Warwick, Universidad de la Plata and the Workshop on Actions, Contributions and Games for useful suggestions.}}}

\author{\begin{tabular}{cc}
Manuel Bagues & Velichka Dimitrova
 \end{tabular}}

\date{\today}

\begin{document}

\maketitle

\begin{abstract}
\noindent We quantify the impact of COVID-19 vaccination on psychological well-being using information from a large-scale panel survey representative of the UK population. Exploiting exogenous variation in the timing of vaccinations, we find that vaccination increases psychological well-being (GHQ-12) by 0.12 standard deviation, compensating for around one half of the overall decrease caused by the pandemic. This effect persists for at least two months, and it is associated with a decrease in the perceived likelihood of contracting COVID-19 and higher engagement in social activities. The improvement is 1.5 times larger for mentally distressed individuals, supporting the prioritization of this group in vaccination roll-outs.

\end{abstract}

\vspace{2cm}

\bigskip 
\vspace{1cm}

\newpage

\doublespacing

\pagenumbering{arabic}
\setcounter{page}{1}
 
\pagebreak

\section{Introduction}
The COVID-19 pandemic and the measures to control the spread of the virus have delivered a profound shock to our lives. Just in the UK, more than 140,000 lives had been lost by October 2021 and, beyond the loss of human life, the pandemic has had also a strong negative impact on the psychological well-being of the population  \citep{pierce2020mental, banks2020mental, proto2021covid}.
Compared to pre-pandemic levels, by November 2020 the average psychological well-being had declined in the UK by 27\% standard deviations and the number of people exhibiting mental distress increased by about 7 percentage points, relative to a baseline of 20\%.\footnote{Authors' calculations using information from \emph{Understanding Society}, 12-item General Health Questionnaire. See section \ref{sec:ghq} for details on how the average psychological well-being and mental distress are calculated.} The decline in mental well-being has been particularly acute among women and young adults \citep{adams2020impact, etheridge2020gender, giuntella2021lifestyle, stantcheva2021inequalities, santomauro2021global}.

Thanks to the unprecedented speed at which science reacted, by the end of 2020 several vaccines against the novel coronavirus had been developed and cleared by regulators around the world. The first mass vaccination program started in December 2020 in the UK. Given the limited supplies of vaccines and logistical constraints, two main population groups were prioritized: i) front-line health and social workers in order to reduce the spread of the virus and protect health systems and ii) elderly and clinically extremely vulnerable populations as they have the greatest mortality risk. Clinical trials have shown that vaccines decrease the probability of contracting and transmitting COVID-19, as well as the severity of infections \citep{abu2021effectiveness,harder2021efficacy, lipsitch2021interpreting, pritchard2021impact, salo2021indirect, thompson2021interim}. However, since in clinical trials participants are unaware of whether they are receiving the vaccine or a placebo, little is known about how vaccination may contribute to the recovery of mental health.

In this paper we exploit evidence from the vaccination roll-out in the UK to estimate the short-term direct impact of vaccination on psychological well-being, with a particular focus on its effect on individuals who were mentally distressed prior to vaccination. This group, which accounted for around 27\% of the UK population, was not explicitly prioritized in the roll-out, despite the dramatic impact that the pandemic had on psychological well-being.
We use data from the UK Household Longitudinal Study \emph{Understanding Society}, a large-scale panel survey representative of the UK population with detailed information on the vaccination status of participants as well as their psychological well-being, assessed through the 12-item General Health Questionnaire (GHQ-12).\footnote{University of Essex, Institute for Social and Economic Research. (2021). Understanding Society: COVID-19 Study, 2020-2021. [data collection]. 9th Edition. UK Data Service. SN: 8644, 10.5255/UKDA‐SN‐8644‐9.}
Taking advantage of the panel structure of the data, we compare the evolution over time in the psychological well-being of individuals who were vaccinated a few weeks before the survey with individuals of the same age and priority group who had not been vaccinated yet. To limit endogeneity concerns, we use invitations for vaccination as an instrument for vaccination. 
We focus mainly on individuals aged 40 to 80 who are not health or social workers, as for this group we can exploit plausibly exogenous variation in the timing of invitations.\footnote{As we explain in more detail below, practically all individuals above 80 years old had received their first jab by the time of the January 2021 survey, as well as health and social workers. On the contrary, relatively few individuals below 40 because had been vaccinated at the time of the last survey wave available (March 2021).}

The first jab improves the mean symptom score in the GHQ-12 by 12\% of a standard deviation (standard error=4\%) and it decreases the number of mentally distressed individuals by 4.3 percentage points (s.e.=2.1).\footnote{In what follows, for simplicity, we abbreviate `point estimate' as `p.e.', `standard error as `s.e.', `standard deviation' as `st. dev.' and `percentage point' as `p.p.'.}
The improvement in psychological well-being provided by vaccination compensates for around one half of the overall decrease in well-being caused by the pandemic.
The impact is 2.5 times as large for individuals with clinically significant levels of mental distress before vaccination. Vaccination increases their average psychological well-being by 29\% st. dev. (s.e.=9\%) and the probability of being mentally distressed decreases by 13 p.p. (s.e.=4.4), compared to individuals with similar levels of distress who had not been vaccinated yet. Interestingly, vaccination does not improve the well-being of individuals with prior concerns about potential unknown side-effects of vaccination (p.e.=-13\% st. dev., s.e.=14\%), who account for around 8\% of the population. The effect of vaccination is slightly larger for women, non-whites, individuals not living with a partner, the less educated, not working from home, not employed and parents of children below 15, but none of these differences are significant at standard levels. The evidence also suggests that the positive impact of vaccines lasts for at least two months, which is the maximum period for which we can observe vaccinated individuals in the survey following their vaccination. Unfortunately, we lack statistical power to estimate the impact of the second jab, as only a few hundred people had received their second dose at the time of the last survey wave. The point estimate is close to zero, but the estimation is too imprecise to discard that the effect was similar to the first jab (p.e.=2\% st. dev., s.e.=8\%). 

To investigate the mechanisms through which vaccination affects psychological well-being, we exploit the rich set of questions available in the survey. Individuals benefit from vaccination through at least two channels. First, vaccinated individuals experience a large decrease in their self-reported expected probability of contracting COVID-19 during the following month (p.e.=-0.17\% st. dev., s.e.= 6\%).
Second, vaccination improves social life and increases the enjoyment of daily activities. Vaccinated individuals are less likely to report that they feel lonely (p.e.=-8\% st. dev., s.e.= 4\%), they are more likely to go for a walk (p.e.=15\% st. dev., s.e.= 6\%) and they are more likely to report that they enjoy normal day-to-day activities (p.e.=15\% st. dev., s.e.= 6\%).
In the case of mentally distressed individuals, the decrease in the expected probability of contracting COVID-19 is similar, but their social life experiences a stronger recovery. 
We do not observe any significant impact on self-reported general health or on the probability of contracting COVID-19. The lack of impact on these dimensions is consistent with vaccination requiring several weeks to be effective in preventing infections, while our analysis focuses on individuals who were vaccinated only a few weeks before the survey. Similarly, there is no significant effect on economic outcomes, probably due also to the short treatment window considered.

The consistency of these results relies on the assumption that the differential timing of invitations is driven by exogenous differences in the speed of the roll-out across different geographical areas, and it is not correlated with time-variant unobserved factors that affect psychological well-being.
A number of robustness tests support this hypothesis. The event study analysis shows that the psychological well-being of vaccinated individuals had evolved in the past similarly to non-vaccinated ones of similar age and priority group. Furthermore, we show that the timing of vaccinations is not correlated with local COVID-19 incidence. Results are also robust when we allow for time-variant shocks at the regional level, and to a number of alternative sampling restrictions and weighting strategies.

Our estimates are larger but less precise when we exploit the regression discontinuity design observed in the March 2021 survey around the 49 year old threshold. Just above this threshold around 90\% of individuals are vaccinated (or have been invited for vaccination) at the time of the survey, compared to only 45\% just below. The well-being of individuals above this age threshold is 13\% st. dev. higher (s.e.=7\%) and the probability of being mentally distressed is 10 p.p. lower (s.e.=4), a significant gap that is not observed in previous waves at this age threshold. The magnitude of the gap is larger when we restrict the analysis to individuals who were mentally distressed in November 2020. Above the 49 years threshold, psychological well-being is higher (p.e.=41\%, s.e.=24) and the likelihood of being mentally distressed is lower (p.e.=-27 p.p., s.e.=11), as measured in March 2021.

Our paper contributes to a growing literature analyzing how the pandemic has affected psychological well-being \citep{adams2020impact, banks2020mental, belot2020unequal, santomauro2021global, etheridge2020gender, giuntella2021lifestyle, pierce2020mental, proto2021covid, proto2021covidbigfive, stantcheva2021inequalities}, and is also related to the strand of literature studying the impact of vaccination on physical health outcomes \citep{abu2021effectiveness, leshem2021population, harder2021efficacy, polack2020safety, pritchard2021impact, thompson2021interim}. 
We contribute to this literature by providing, up to the best of our knowledge, the first evidence on the causal impact of vaccination on psychological well-being. 
Our work also speaks to the discussion on the optimal design of vaccination roll-outs and the selection of priority groups \citep{buckner2021dynamic, mazereel2021covid, Smith2021covid, stip2021people, yang2021should}. Our findings suggest that, in addition to physical health, mental health should be also taken into account.

\section{Institutional Context} \label{sec:institutional_context}

In early December 2020, the UK became the first country to approve a COVID-19 vaccine for emergency use and to start a mass vaccination program.
The vaccination roll-out was conducted by the National Health Service (NHS) according to priority groups. The ordering was mainly based on age as the available evidence indicated that it was the single greatest risk for mortality from COVID-19 \citep{jcvi-prioritygroups}.
Priority groups also included residents and staff of care homes, front-line health and social care workers, clinically extremely vulnerable, and at-risk groups with underlying health conditions (see Figure \ref{fig:priority-poster}). The vaccination roll-out did not explicitly target individuals with low psychological well-being, although individuals with extreme psychiatric conditions were considered clinically vulnerable and included as a priority group 6, following the vaccination of all individuals aged 65 and over \citep{jcvi-prioritygroups, smithbmjopinion}.
By April 1st 2021, 59\% of its 18+ population had received their first jab.\footnote{Ratio between the total number of vaccinations given to people of all ages and the mid-year 2020 population estimate for people aged 18 and over. Source: Public Health England. Available at https://coronavirus.data.gov.uk.}

The NHS sent invitations to registered patients according to the priority ordering \citep{delivery-plan}, and individuals receiving an invitation could schedule an appointment for vaccination, which was implemented mainly by local GPs. The system did not allow to get an appointment without an invitation letter or to be vaccinated without an appointment.

The beginning of the vaccination process in December 2020 coincided in time with a new wave of the pandemic, which led to the introduction of the third national lockdown on 6 January 2021. The lockdown rules were uniform across different areas of the country. The population was requested to stay at home and they were not allowed to meet indoors with people from other households. 
Following the decline in the infection rate, on February 22 a roadmap for lifting the lockdown was published \citep{out-lockdown-guidance}. The first phase started on March 8, and it included the return to face-to-face education in schools and the relaxation of social contact restrictions.
The `stay at home rule' was removed on March 29.

\section{Data} \label{sec:data}

We use data from the UK Household Longitudinal Study \emph{Understanding Society},\footnote{University of Essex, Institute for Social and Economic Research. (2021). Understanding Society: COVID-19 Study, 2020-2021. [data collection]. 10th Edition. UK Data Service. SN: 8644, http://doi.org/10.5255/UKDA-SN-8644-10.} a large-scale longitudinal survey which relies on probability samples representative of the UK population \citep{benzeval2021high}. 
We focus on the eight COVID survey waves that were conducted between April 2020 and March 2021. These waves were conducted mainly online, with a telephone follow-up of web non-respondents who reside in households where no one regularly uses internet for some waves. The survey population includes around 20k individuals, of which 42\% participated in all eight waves, 52\% participated in all but one wave, and 58\% in all but two waves. 

Table \ref{table:summary}, columns 1-4, provides information on the main summary statistics for the adult population in the survey. Around 53\% are women, the average age is 51, 90\% self-identify as white, 90\% were born in the UK, 75\% live in an urban place, 29\% are college graduates (with a degree qualification), 63\% live with a partner and 22\% are parents of a child who is less than 15 years of age.\footnote{All descriptive statistics and analysis are performed using the weights, stratification and clustering defined by \emph{Understanding Society}.} More detailed information about the definition of each variable is available in Table \ref{table:variable_definitions}.

\subsection{Vaccination process}\label{sec}

Two survey waves took place during the vaccination process. Wave 7 was launched on 27 January 2021, when the vaccination roll-out was reaching individuals aged around 60 to 80 years old, and lockdown rules were strictest. Wave 8 was launched on 24 March 2021, during the first phase of the lockdown easing and when individuals around 40 to 60 years old were being vaccinated.

According to the survey, by the end of January 2021 around 24\% of UK adults had received an invitation for vaccination, of which 18\% had already received the 1st jab and 4\% had made an appointment for vaccination (see Figure \ref{fig:vax-status}). By the end of March 2021, when the following survey wave took place, the share of individuals who had received an invitation had increased to 71\%, of which 65\% were vaccinated, 3\% had an appointment, and 3\% had not booked an appointment yet. Around 5\% of individuals had already received also their second jab. 

\subsection{Priority groups}

The survey provides detailed information allowing to identify groups that were prioritized in the roll-out. Following the categorization created by the NHS, it classifies respondents into (i) individuals with no risk of serious illness if they contract COVID-19 (61\% of the population), (ii) clinically vulnerable (36\%) and (iii) clinically extremely vulnerable (5\%).\footnote{This classification is based on a two-level categorization created by the NHS in May 2020 relying on underlying disease or health condition. More detailed information is available in the User Guide for \emph{Understanding Society} COVID-19, page 33 \citep{user-guide}. As pointed out in the User Guide for Understanding Society COVID-19, over time the NHS has refined its definitions, and this classification is likely to be broader than more precise NHS records. This variable therefore is likely to be an overestimation of those at risk of serious illness if they are infected with COVID-19.}
Approximately 10\% of the survey population had also received a letter, text or email from the NHS or Chief Medical Officer indicating that they are at risk of severe illness if they catch coronavirus, and they are advised to stay at home (`shielded').
Furthermore, the roll-out prioritized health and social workers, who account for around 12\% of the survey population, and individuals who care for someone who is sick, disabled or elderly, close to 9\%.\footnote{We identify health or social workers using the information provided by the survey on the industry of main activity ("Human Health and Social Work Activities" group in variable \textit{jbindustry}) and on the key work sector ("Health and social care" group in variable \textit{keyworksector}).}

\subsection{Psychological well-being: general health questionnaire}
\label{sec:ghq}
The survey measures psychological well-being using the General Health Questionnaire (GHQ-12). This questionnaire includes 12 questions and it is commonly used as a screening tool to detect current state mental disturbances and disorders in primary care setting \citep{goldberg1979scaled}. The specific questions are available in Appendix \ref{sec:app-ghq}. Each questions has four possible answers: ``not at all", ``no more than usual", ``rather more than usual" and ``much more than usual".
We rely on two measures derived from the GHQ-12 index that have been broadly used in the literature \citep{pierce2020mental, banks2020mental, proto2021covid, etheridge2020gender}. First, we measure psychological well-being using the \textit{mean symptom score}, which is constructed summing up the 12 GHQ items, coded to run from 0 to 3, resulting in a scale from 0 (the least distressed) to 36 (the most distressed). To ease interpretation we standardize it and invert it, so that higher values indicate higher psychological well-being. Second, we rely on a \textit{caseness indicator} that captures more acute cases of mental distress \citep{morris2017health}. This indicator takes value one for individuals who have replied ``rather more than usual" and ``much more than usual" in at least four of the twelve questions of the GHQ-12.

As shown in Figure \ref{fig:timelines}, during the first months of the pandemic, mental health measures deteriorated dramatically. In April 2020 the average psychological well-being was 21\% st. dev. lower than in 2019. As cases fell during the summer, mental health improved, but with the arrival of the second wave in Fall 2020, the average psychological well-being dipped even further. By November 2020 it had declined by 27\% st. dev. compared to pre-pandemic levels. Similarly, in Fall 2020 about 26\% of the respondents showed clinically significant levels of mental distress, as measured by the caseness indicator, compared to just under 20\% in the years leading up to the pandemic.

Psychological well-being exhibits a strong age-gradient. The mean symptom score is around 25\% st. dev. higher among individuals above 60 years old than for individuals between 40 and 60. Psychological well-being is also correlated with physical health, but the relationship is relatively weak. Among individuals who are extremely clinically vulnerable, 35\% of them are also mentally distressed, compared to 27\% in the rest of the population.

\subsection{Physical health and COVID-19}

The general health of individuals is reported on a Likert scale. Around 48\% consider that their health is excellent or very good, and only 4\% report having poor health. The survey also provides information about the incidence of COVID-19. In each survey wave about 5\% of respondents report having experienced symptoms since the previous wave that could have been be caused by coronavirus, and 2\% report that they have tested positive in a COVID-19 test. 

Survey respondents also report their expected likelihood of contracting COVID-19. In November 2020, prior to the vaccination roll-out, around 9\% of respondents declare that it is likely or very likely that they will get infected during the following month (see Figure \ref{fig:covid-risk} (a)). Interestingly, the self-reported risk of contracting COVID-19 has a strong correlation with the actual probability of reporting a positive test result in the following survey wave.
Around 4.3\% of individuals who report a very high COVID-19 risk assessment test positive during the following two months, compared to 1.8\% of individuals who report a very low risk (see Figure \ref{fig:covid-risk} (b)). The perceived risk decreases monotonically with age. For instance, around 11\% of individuals aged between 40 and 60 consider that it is likely or very likely that they will contract COVID-19 during the following month, compared to only 4\% of individuals above 60 years old.

\subsection{Attitude towards vaccines}
\label{sec:vax-attitudes}

Survey respondents declare that their main motivations for getting the jab are to avoid catching COVID-19 (55\%), to recover social life (10\%) and to protect other people (9\%) (see Figure \ref{fig:vax-reasons}). On the other hand, the main concern is the possible existence of unknown side effects (11\% of respondents).

In November 2020, before the beginning of the roll-out, around 18\% of individuals report that they are unlikely or very unlikely to vaccinate (See Figure \ref{fig:vax-attitudes}). The share of people unwilling to vaccinate decreases over time as the roll-out is implemented, first to 8\% in January 2021 and then to 5\% in March 2021.

\subsection{Social life and enjoyment of daily activities}

We use three proxies for the intensity of social interactions.
First, individuals are asked how often they felt lonely in the previous 4 weeks. In November 2020, around 44\% of people report feeling lonely some of the time or often, compared to 40\% in 2019 before the pandemic. Self-reported loneliness tends to decrease with age. About 56\% of individuals below 40 report feeling lonely sometimes, compared to 36\% of middle-age individuals and 30\% of individuals above 60. Second, in some survey waves, including in January 2021, individuals report how many days they walk weekly. The average individual in the sample walks five days a week for at least 10 minutes.
Third, as part of the GHQ-12 questionnaire, individuals are asked whether they enjoy day-to-day activities. Around 55\% reply than they do so less than usual, compared to only 5\% who enjoy them more.

\subsection{Labor market and household finances}
Around 60\% of individuals are employed, they work on average 28 hours a week and 36\% work from home always or often. Household weekly income is close to GBP 650 (USD 900).
Approximately 2/3 or respondents report that they are living comfortably or doing alright, and most of them (78\%) do not expect their financial situation to change in the following three months. During the past 4 weeks they have saved on average around GBP 250 (USD 350). To measure the marginal propensity to consume, respondents are asked how receiving a (hypothetical) one time payment of GBP 500 (USD 670) would affect their spending, borrowing and saving behaviour over the following three months. Around 19\% of them declare that they would spend more.\footnote{We combine the responses to a question where the origin of the payment if not specified and a question where the fictitious payment was done by the government (variables `mpc1' and `mpc1b').}

\subsection{Other data sources: COVID-19 incidence and vaccination rates}
We complement the information provided by the survey with information on COVID-19 incidence at the local level, as measured by the daily number of COVID-19 positive tests per 100,000 inhabitants in the corresponding Middle Layer Super Output Areas (MSOAs). The average MSOA includes around 4,000 households.

We also collected administrative data from Public Health England on the percentage of people by age group who received the 1st COVID-19 vaccination over time. As shown in Figure \ref{fig:uptake-heatmap}, the progression of the vaccination uptake followed within age brackets. Once the roll-out reached a cohort, it took just a few weeks to complete its vaccination. For instance, individuals aged 75-79 started to be vaccinated in mid-January and by mid-February their vaccination had been completed. 

\section{Empirical Analysis} \label{sec:empirical_analysis}

\subsection{Factors predicting invitations and vaccinations}
\label{sec:determinants_vaccination}
\begin{figure}[htbp]
\caption{Vaccination rate by age.}
\begin{center}
\subfloat[January 2021 (wave 7)]{\includegraphics[height=2.3in]{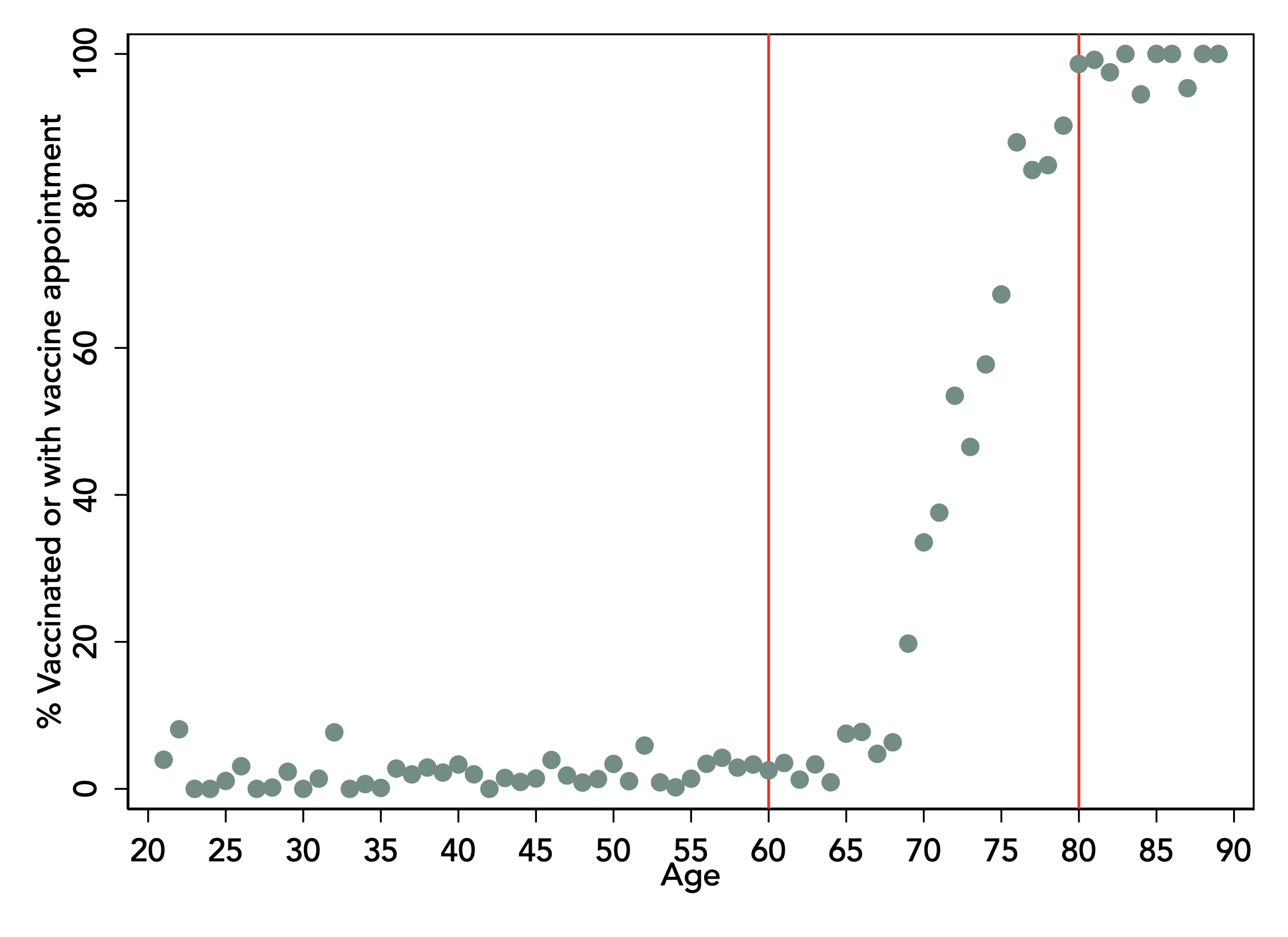}}
\subfloat[March 2021 (wave 8)]{\includegraphics[height=2.3in]{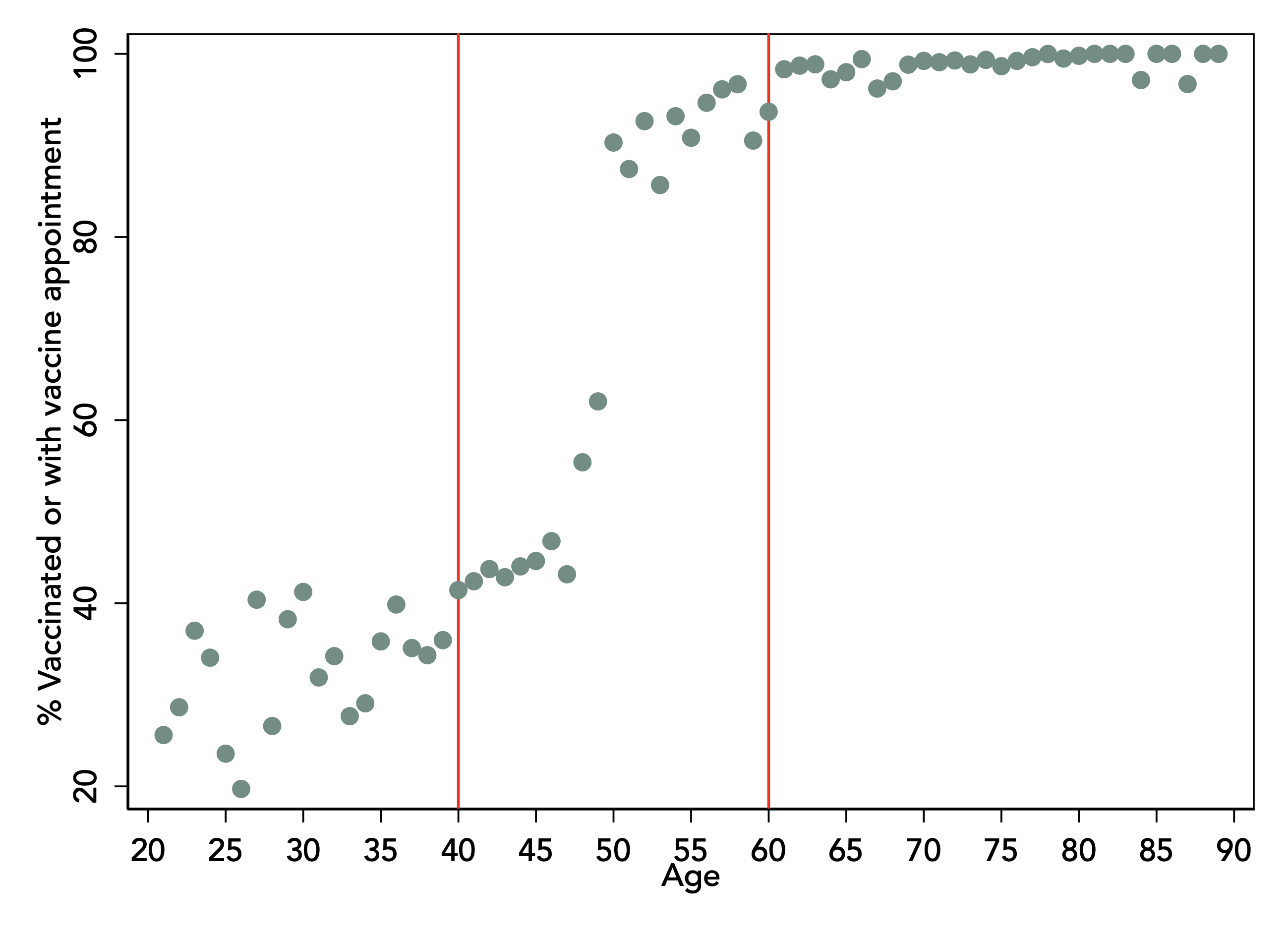}}
\end{center}
\vskip 2mm
\caption*{\textit{Notes:} Percentage of individuals who were vaccinated or had an appointment for vaccination by age group in wave 7 (panel a) and wave 8 (panel b). Source: Authors' elaboration using data from \emph{Understanding Society}.}
 \label{fig:vaccination_by_age}
\end{figure}

We start by examining the main factors that predict whether an individual in the survey has been invited for vaccination. As expected, invitations for vaccination exhibit a strong age gradient (see Figure \ref{fig:vaccination_by_age}). In the January 2021 survey, above 99\% of individuals aged 80 or older had already received an invitation, compared to only 20\% of younger adults. Similarly, at the time of the March 2021 survey, invitations had been sent to 99\% of individuals above 50, and to only 40\% of younger individuals.

To investigate the role played by other individual characteristics, we estimate the following equation:

\begin{equation} \label{eq:invitation_vaccination}
\begin{split}
y_{it} &= \sum_{a=41}^{80} \alpha_a \cdot I(a=age_{i}) + \beta \cdot I(t=March_{2021}) + X_{it} \cdot \gamma + \epsilon_{it} 
\end{split}
\end{equation}

where $y_{it}$ is an indicator variable that takes value one if individual $i$ had been invited for vaccination at the time of survey wave $t$. As explanatory variables we consider a full set of age dummies, a survey wave indicator and a vector of individual characteristics which includes information on prior mental and physical health, industry, occupation and keyworker status, and the (standardized) local COVID-19 incidence rate measured at the beginning of the roll-out. We use survey weights in our estimation and we cluster standard errors by survey strata.
We restrict the analysis to individuals aged 61-80 in the January 2021 survey wave and 40-60 in the March survey wave, as only these age groups experience some relevant (within age) variation in vaccination status during this period. Our results are essentially unchanged if we consider instead the overall sample.

The estimates are reported in Table \ref{table:determinants}, column (1). As expected, members of priority groups more likely to have been invited for vaccination. The main effect is observed for health and social care keyworkers (24 p.p).  We also observe significant effects for other priority groups: carers (11 p.p.), clinically extremely vulnerable individuals (6 p.p.) and vulnerable individuals (5 p.p.), and individuals who were shielding (10 p.p.). 
On the other hand, the survey confirms that individuals with high levels of mental distress were not prioritized (p.e.=-0.009, s.e.=0.015). 

Most importantly for our empirical strategy, we do not observe any correlation between the speed of the roll-out and the local incidence of COVID-19 at the beginning of the roll-out. The estimate is a rather precise zero (p.e.=-0.003, s.e.=0.006), indicating that in areas where COVID-19 incidence is one standard deviation higher the probability of receiving an invitation may be up to 1.5 p.p. higher or 0.9 p.p. lower.

\paragraph{Vaccination}

The immense majority of people who had received an invitation, around 95\%, got vaccinated or was was waiting for an appointment. To study which factors affected the decision of getting vaccinated, we estimate equation (\ref{eq:invitation_vaccination}) for the sample of individuals who received an invitation, taking as outcome variable taking a dummy that takes value one if the individuals have made an appointment or received their first vaccine dose, and zero if individuals received an invitation but failed to make an appointment.

As shown in Table \ref{table:determinants}, columns (2), vaccination rates are significantly higher among priority groups and among individuals living with a partner. Individuals' concerned with the potential existence of vaccination side effects are 6 p.p. less likely to vaccinate. We observe also some differences by geographic area, but we do not observe any significant correlation between vaccination rates and COVID-19 incidence. Similarly, we do not observe that previous mental distress in any way predicts vaccination.

\subsection{Short-term impact of vaccination} \label{sec:short-term}

In this section we investigate the impact of vaccination on psychological well-being during the first weeks after vaccination. First, we use a difference-in-differences (DID) empirical strategy. We compare individuals vaccinated in the weeks leading up to the survey with individuals of the same age and priority group who had not been vaccinated yet. Second, to address potential endogeneity concerns and reverse causality, we combine the DID estimation with an instrumental variables approach, where we use as an instrument for vaccination whether individuals have received an invitation. Third, we study a number of mechanisms that may underlie the observed relationship between vaccination and psychological well-being. Fourth, we examine whether the effect varies depending on the prior level of psychological well-being. Fifth, we verify the robustness of our estimates to alternative specifications, different sample restrictions, and we test for the existence of non-random participation into the survey. Sixth, we consider several extensions and heterogeneity analyses. Finally, we estimate the short-term impact of vaccination using a regression discontinuity design, exploiting a sharp discontinuity around age 49 in the vaccination rate of individuals surveyed in March 2021.

\subsubsection{Difference-in-differences} \label{sec:did}

We estimate the short-term impact of vaccination using a stacked DID model with two treatment groups: (i) individuals aged 61-80 who were vaccinated at the time of the January 2021 survey wave and (ii) individuals aged 40-60 who were vaccinated in the March 2021 survey wave.\footnote{A recent literature has pointed out that in DID setups with staggered adoption the interpretation of standard two-way fixed effects estimates as the average treatment effects for the treated sub-populations may not be adequate in the presence of heterogeneity and dynamic effects, even when the `parallel trends assumption' holds (e.g. see \cite{callaway2020difference}). The use of a stacked DID helps to address these concerns by ensuring that (i) we are comparing `switchers' only to `not yet treated' individuals and (ii) both groups are receiving the appropriate weight. Note also that dynamic effects are not present here, as we only consider the first period in which individuals are treated.}
The control groups include other individuals of the same age and priority group who were not vaccinated by the time of the corresponding survey. For both age groups, we consider one survey wave after the treatment and six survey waves before, i.e. we consider individuals aged 61-80 from April 2020 until January 2021 (waves 1-7 of the survey) and individuals aged 40-60 from May 2020 until March 2021 (waves 2-8 of the survey).
As shown in Figure \ref{fig:uptake_by_age}, individuals in the treatment group were vaccinated mostly during the four weeks prior to the survey, while individuals in the control group were vaccinated in the following weeks, implying that we are comparing individuals who were vaccinated a few weeks before the survey with individuals of the same age and priority group who had not been vaccinated yet, but would be vaccinated in a few weeks.

More specifically, we estimate the following equation:

\begin{equation} \label{eq:did}
\begin{split}
y_{it} &= \sum_{j\in \text{sample}} \alpha_j \cdot I(j=i) + \sum_{e=-6}^1 \sum_{a=41}^{80} \beta_{e,a} \cdot I(e=t) \cdot I(a=age_{i}) + \\ 
&+ \sum_{e=-6}^1 \sum_{p} \sum_{g} \lambda_{e,p,g} \cdot I(e=t) \cdot I(p=prioritygroup_{it}) \cdot I(g=agegroup_{it}) + \gamma \cdot \text{Vaccinated}_{it} + \epsilon_{it} 
\end{split}
\end{equation}

where we have denoted $t=1$ for the survey wave when a given age group could have been first vaccinated (i.e. for individuals aged 61-80, $t=1$ refers to January 2021 and, for individuals aged 40-60, $t=1$ in March 2021), and $y_{it}$ is the standardized GHQ-12 score for individual $i$ at time $t$. We include as controls a set of individual fixed effects (first term on the right-hand side), a set of event time fixed effects interacted with age dummies (second term), and a set of event time fixed effects interacted with age group (40-60 or 61-80) and priority group (third term).\footnote{In particular, we consider the following priority groups: keyworkers, `shielded', clinically vulnerable, extremely clinically vulnerable, and carers.} $\text{Vaccinated}_{it}$ is a dummy variable that takes value one if individual $i$ was vaccinated or had arranged an appointment for vaccination at time $t$.\footnote{Around 3\% of the population had an appointment for vaccination at the time of the survey but was not vaccinated yet. We include them in the treatment group to facilitate that our DID results can be comparable to the IV analysis, where in order to satisfy the exclusion restriction, we need to group together individuals vaccinated and with an appointment. Otherwise, if we assigned zeroes to individuals with an appointment in a context where having an appointment affects psychological well-being, that would violate the exclusion restriction of the IV analysis, which requires that invitations only affect psychological well-being through its effect on the instrumented variable. As shown in the robustness section, this choice does not affect results.} In all estimations, we use survey weights and we cluster standard errors by survey strata.

In the above equation, individual fixed effects capture unobserved time-invariant heterogeneity. Furthermore, the inclusion of wave fixed effects interacted with age allows to control for the potential existence of time-variant age-specific shocks. This is likely to be a relevant concern in this context, as we observe that the psychological well-being of different age groups had evolved differently in the past. For instance, in the summer of 2020, individuals above 60 experienced a significant improvement in their psychological well-being, compared to younger individuals, perhaps due to the differential impact of the pandemic and lock-downs on the well-being of different age groups (Figure \ref{fig:trend_by_age}). The third set of fixed effects allows for the existence of time-variant shocks that may affect differently groups of individuals in different priority groups. Given these sets of controls, the identification of $\gamma$ relies on the the timing of vaccination being exogenous for individuals of the same age and priority group. 

To estimate equation (\ref{eq:did}), we exclude health and social workers as they were strongly prioritized and mostly vaccinated by the end of January 2021. For the same reason, we also exclude from the analysis 158 individuals aged 40-60 (2\% of the sample) who had been invited for vaccination at the time of the January 2021 survey, as they are likely to have been targeted as members of some special group. 
The remaining sample includes around 48k observation corresponding to 8k different respondents. As shown in section \ref{sec:robustness}, these sample restrictions do not affect our results in any relevant way.

In Table \ref{table:maintable}, we provide in column (1) the results from a simple OLS regression where we only control for survey wave fixed effects. The average psychological well-being of vaccinated individuals is 21\% st. dev. higher (s.e.=4\%) compared to non-vaccinated individuals in the same survey wave. Naturally, this estimate is likely to reflect a combination of selection biases and causal effects. In column (2), we include individual fixed effects, and the estimate becomes smaller: p.e.=4.7\%(s.e.=2.4\%), indicating that vaccinated individuals are positively selected in terms of their underlying psychological well-being, partly reflecting that they tend to be older and psychological well-being is increasing with age.
In column (3) we allow for time-variant age-specific shocks, which leads to a substantial increase in the magnitude of the estimate: p.e.=13\% (s.e.= 3\%). A possible explanation for the change in the magnitude of the estimate is that individuals above 60 experienced between November 2020 and January 2021 relative to other age groups, perhaps due to their larger sensitivity to lockdown measures and increases in the infection rate. This age group was more likely to be vaccinated during this period, and not accounting for this age-specific negative shock biased downwards the estimate of the impact of vaccination.
Finally, in column (4) we report the estimate with the full set of controls included in equation (\ref{eq:did}), allowing also for time-variant shocks affecting different priority groups. This additional set of controls leaves the estimate unchanged: p.e.=12\% (s.e.=3\%).

\paragraph{Event study}

The validity of the DID estimates relies on the assumption that, in the absence of the treatment, both groups would have evolved similarly. To explore the plausibility of this hypothesis we conduct an event study analysis and we estimate the following equation:

\begin{equation} \label{eq:eventstudy}
\begin{split}
\text{y}_{it} &= \sum_{j\in \text{sample}} \alpha_j \cdot I(j=i) + \sum_{e=-6}^{1} \sum_{a=41}^{80} \beta_{w,a} \cdot I(e=t) \cdot I(a=age_{i}) + \\\
&+ \sum_{\substack{e=-6 \\ e\neq 0}}^1 \gamma_k \cdot I(e=t) \cdot treatmentgroup_{i} + \epsilon_{ite} 
\end{split}	
\end{equation}

where $y_{it}$ is the psychological well-being of individual $i$ at event time $t$. We control for individual fixed effects (first term on the right-hand side), event time fixed effects interacted with age (second term), and a full set of event time dummmies interacted with the treatment group identifier (third term). The variable $treatmentgroup_{i}$ takes value one for individuals aged 61-80 who were vaccinated at the time of the January 2021 survey and individuals aged 40-60 who were vaccinated at the time of the March 2021 survey. We omit the event time dummy at t = 0, implying that the event time coefficients measure the impact relative to the survey wave just before vaccination.

\begin{figure}[htbp]
\caption{Event study - Impact of vaccinations on psychological well-being}
\begin{center}
\includegraphics[height=2.5in]{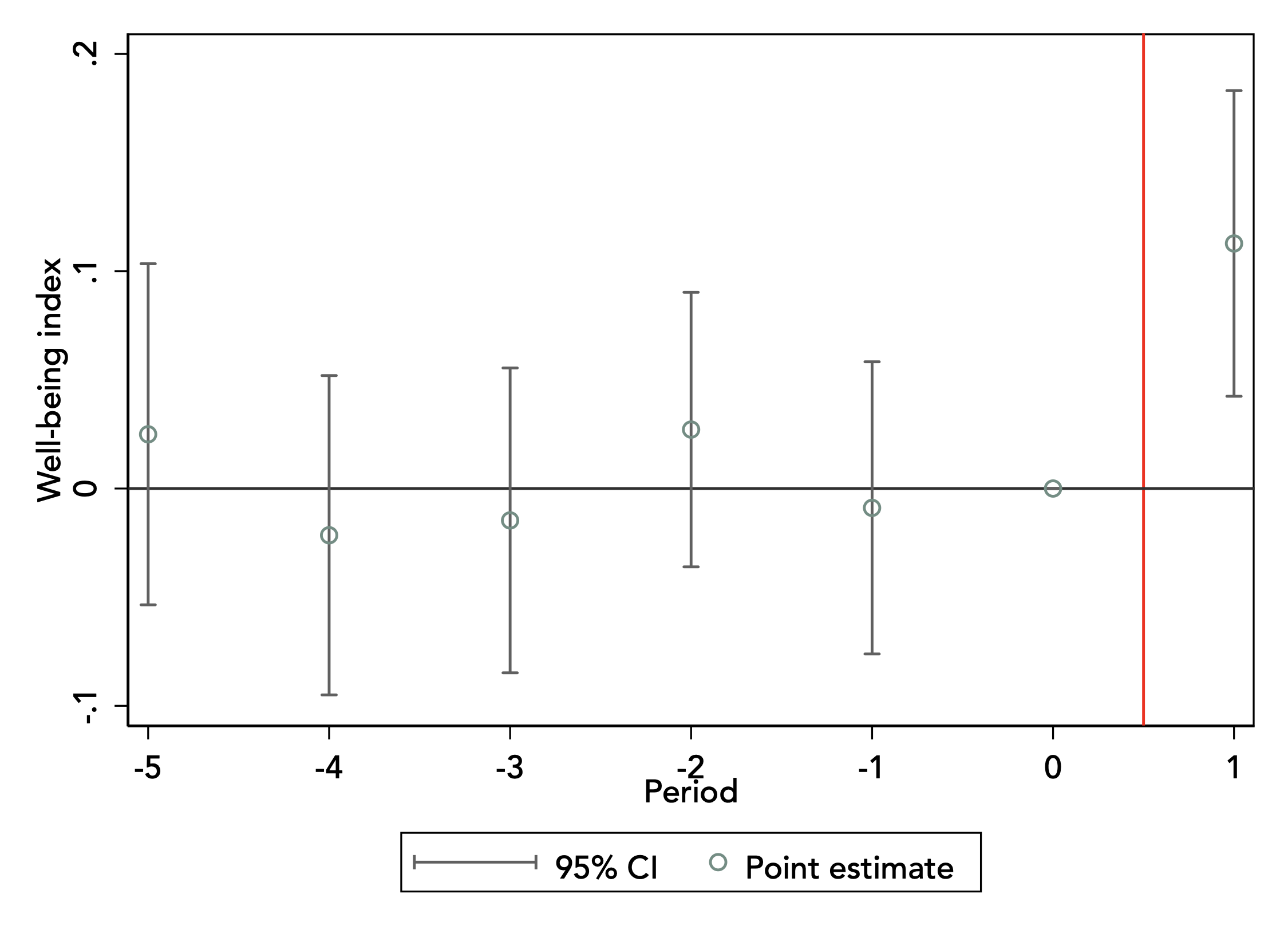} \\
\end{center}
\caption*{\textit{Notes:} The figure reports the estimates from the event study analysis of the DID (see equation \ref{eq:eventstudy}). The outcome variable is the standardized and inverted GHQ-12 mean score. The treatment group includes individuals aged 61-80 who were vaccinated at the time of the January 2021 survey and individuals aged 40-60 who were vaccinated in March 2021 survey. For individuals age 61-80 (40-60), we denote the November 2020 (January 2021) survey wave as the baseline period, and index all waves relative to that one.}
 \label{fig:event_study_vaccination}
\end{figure}

As shown in Figure \ref{fig:event_study_vaccination}, both groups followed parallel trends in the past, supporting the assumption that they would have also followed parallel trends in the absence of the treatment. Moreover, consistent with our previous results, the event study also shows that psychological well-being increases significantly following vaccination.

\subsubsection{Instrumental variables} \label{sec:iv}

There are at least two potential threats to the validity of the DID empirical strategy. First, there might a problem of reverse causality. Low psychological well-being may affect vaccination decisions. Furthermore, there might be an omitted variable bias if there are unobserved (time-variant) factors that affect simultaneously the probability of being vaccinated and psychological well-being. To address these concerns, we estimate equation (\ref{eq:did}) using invitations as an instrument for vaccination.

As shown in column (5) of Table \ref{table:maintable}, the IV estimate is slightly less precise but the point estimate is practically identical (p.e.$_{IV}$=12\% st. dev, s.e.=4\%). In columns 6 and 7 we analyse separately the impact of vaccination for individuals aged 61-80 (waves 1-7) and 40-60 (waves 1-8). The impact of vaccination is similar for both groups: p.e.$_{61-80}$=12.1\% vs. p.e.$_{40-60}$=11.9\%.

So far we have used an outcome variable the GHQ-12 mean score. In Table \ref{table:maintable-distress} we reproduce all our analyses using as an outcome variable an indicator for mentally distressed individuals. As shown in column 5, the IV estimate, which is our preferred specification, indicates that vaccination decreases the probability of being mentally distressed by 4.3 p.p. (s.e.=2.1 p.p.)

\paragraph{Exogeneity}

The validity of the IV strategy requires the exogeneity of the instrument, conditional on controls. In this particular context, given that the identification relies on a DID specification, this assumption is equivalent to the standard parallel trends assumption, which requires that the group of individuals who have not been invited yet for vaccination provide a good counterfactual for the evolution of well-being among individuals who have been invited for vaccination. To investigate the plausibility of this assumption, we conduct an event study analysis. We estimate equation (\ref{eq:eventstudy}), using as treatment group individuals that were invited for vaccination. More precisely, the treatment group includes individuals aged 61-80 who had received an invitation for vaccination by the end of January 2021, and individuals aged 40-60 who had received an invitation by the end of March 2021. The control group includes all other individuals of the same age. As shown in Figure \ref{fig:event_study_invitation}, the event study confirms that treated and non-treated individuals evolved similarly in the past, supporting the exogeneity of the instrument.

\paragraph{Exclusion restriction}

In addition to the exogeneity assumption, for the consistency of the IV estimator the exclusion restriction requires that invitations for vaccination only affect psychological well-being through their impact on vaccination. This assumption would not be satisfied if invitations may somehow affect the well-being of individuals who decline to get vaccinated, a possibility that seems unlikely. 

\paragraph{Monotonicity condition}
Finally, the monotonicity condition requires that, for all groups of individuals, receiving an invitation does not decrease the probability of being vaccinated. We do not observe in the data any individuals getting vaccinated without an invitation, indicating that this assumption is also satisfied.

\subsubsection{Mechanisms} \label{sec:mechanisms}

Individuals report in the survey that their main motivations for getting vaccinated are related to health, social life, and getting back to work; and their main concern against vaccination are potential side-effects. Below we explore the role played by each of these factors.

\paragraph{Physical health and COVID-19}
The observed positive impact on mental health might potentially reflect that vaccinated individuals are experiencing an improvement in their current health. In our setup, this possibility is in principle limited, given that individuals received their first jab just a few weeks before the survey and clinical trials indicate that COVID vaccines are ineffective during the first weeks \citep{polack2020safety}. If anything, some subjects might have experienced negative side effects around 1 to 2 days after the vaccine, including pain in the arm, feeling tired, headache, general aches, or mild flu like symptoms \citep{menni2021vaccine}. As shown in Table \ref{table:mechanisms-health}, columns 1-3, vaccination does not seem to affect health in the short-term horizon that we consider here. There is no significant effect on the probability of having symptoms of COVID-19 in the previous month, testing positive, or self-reported general health.

Instead, vaccination affects individuals' expectations of contracting COVID-19 during the following month, which decreases by 17\% st. dev. (s.e.=6\%) (Table \ref{table:mechanisms-health}, column 4)

\paragraph{Social life}

Another mechanisms that seems to mediate the impact of vaccination on psychological well-being is the improvement of social interactions (see columns 5-7). Following vaccination, we observe a decrease in self-reported loneliness of 8\% st. dev. (s.e.= 4\%) and an increase in daily walking of 15\% st. dev., s.e.= 6\% (Table \ref{table:mechanisms-health}. Vaccination also leads to a higher enjoyment of daily activities (15\% st. dev., s.e.= 6\%), which is one of the 12 items of the GHQ-12.

\paragraph{Labor market and household finances}

As shown in Table \ref{table:labour-financial}, we do not observe any significant effect on labor market outcomes, saving behavior or consumption. 
There are at least two possible explanations for this lack of impact.
Individuals were vaccinated just a few weeks before the survey, limiting the scope for job market effects. Moreover, during the period that we study, end of January and end of March 2021, work opportunities were limited by a lockdown that required working from home for a large share of the population.

\paragraph{Attitude towards vaccines}

The fear of future unknown side-effects might mediate the impact of vaccines on psychological well-being.
We estimate our IV-DID model allowing for an interaction between vaccination and a dummy variable that takes value zero for individuals who had expressed concerns with side-effects in the November 2020 survey and value one otherwise.
In column (1) of Table \ref{table:mech-vax} we report the results of the first stage estimation. Most individuals concerned with side-effects get vaccinated when they receive an invitation (p.e.=75 p.p., s.e.=7 p.p.), though their take-up rate is significantly lower than the rest of the population (p.e.=17 p.p., s.e.=7 p.p.).
In column (2) we report estimates for the reduced form analysis (i.e. we regress psychological well-being on invitations). Invitations have no significant impact on the psychological well-being of individuals concerned with side-effects (p.e.=-9\% st. dev., s.e.=10\%), in contrast with the significantly larger impact experienced by other people (p.e.=22\% st. dev., s.e.=11\%). In column (3) we estimate the 2-stage least square model, instrumenting vaccination decisions with invitations. As expected, the point estimate is negative and slightly larger than the reduced-form (p.e.=-13\% st. dev., s.e.=14\%). Overall, our results suggest that vaccine hesitancy undermines the psychological benefits of the vaccination roll-out.

\subsubsection{Impact on mentally distressed individuals} \label{sec:distress}

Our analysis has shown that vaccination has a large and significant positive impact on psychological well-being. Next, we investigate whether this effect is different for the subset of individuals who were mentally distressed before vaccination. We estimate equation (\ref{eq:did}) allowing for an interaction between vaccination and an indicator variable that takes value one for individuals who had clinically significant levels of mental distress in September 2020 (\textit{$Distressed^{Sept 2020}$}), and we use as instruments the variable \textit{invitation} and its interaction with  \textit{$Distressed^{Sept 2020}$}. We restrict this analysis to the last three waves of the survey (November 2020, January 2021 and March 2021), to ensure that the right-hand side measure of mental distress is predetermined relative to the outcome variables.

We report these results in Table \ref{table:mentally-distressed}. As shown in column (2), the positive impact of vaccination is 2.5 times as large for mentally distressed individuals (29\% vs. 12\%). The impact is also significantly stronger when we consider as the outcome variable a dummy for being mentally distressed at time $t$. Individuals who were mentally distressed in September 2020 are 13 p.p. (s.e.=4.4) less likely to remain in this state after vaccination, compared to other mentally distressed individuals individuals who were not vaccinated (column 3).

In columns (4)-(6), we investigate the possible mechanisms.
We do not observe any differential impact on the perceived risk of contracting COVID-19. However, individuals with high levels of initial of mental distress seem to benefit more in terms of their social interactions. They experience a larger decrease in self-reported loneliness and a larger increase in their enjoyment of daily activities.\footnote{We do not consider the variable \textit{weekly walking}, as it is available only in two waves.}

\begin{figure}[htbp]
\caption{Impact of vaccination on psychological well-being, by previous level of mental distress}
\begin{center}
\includegraphics[height=2.5in]{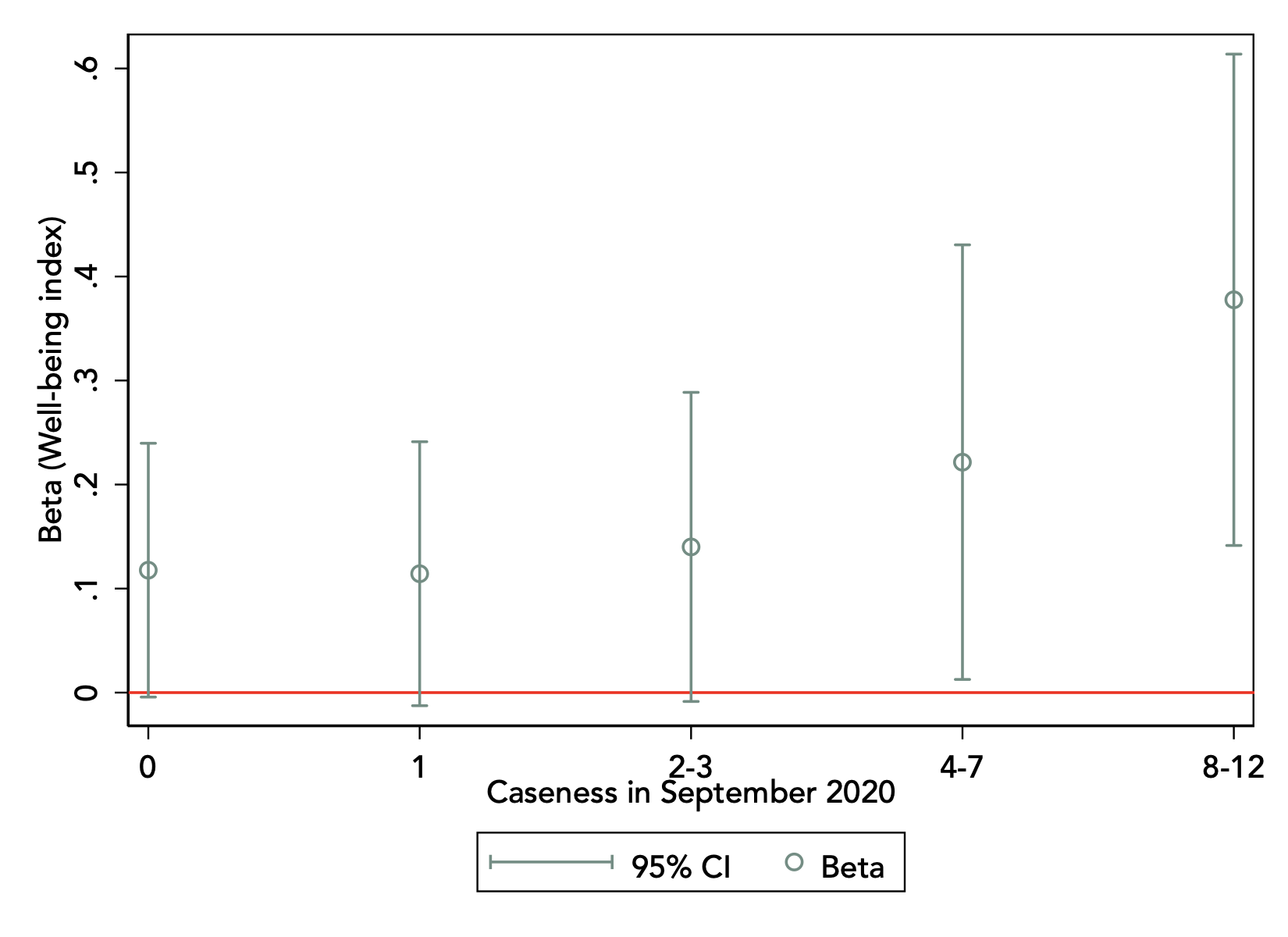} 
\end{center}
\caption*{\textit{Notes:} Coefficient plot of the impact of vaccination on the standardized and inverted GHQ-12 mean score in March 2021, by caseness score measured in September 2020. The caseness score is derived by scoring in the GHQ-12 the "not at all" and "no more than usual" responses as 0, and the "rather more than usual" and "much more than usual" responses as 1, and summing them up, resulting in a range 0-12.}
\label{fig:casenessgroups-wellbeing}
\end{figure}

So far we have followed the convention of relying on an indicator variable which classifies individuals as distressed if they have replied ``rather more than usual" and ``much more than usual" to at least four or more questions of the GHQ-12. To investigate further how the impact of vaccination varies with the level of initial mental distress, we estimate the same equation interacting vaccination with a more granular measure of mental health, the \textit{caseness score}. This score is derived by scoring the ``not at all" and ``no more than usual" responses as 0, and the ``rather more than usual" and ``much more than usual" responses as 1, and summing them up, resulting in a range 0-12. We consider in our regression five possible categories: zero questions (54\% of the sample), one question (14\%), 2 or 3 questions answers (12\%), 4-7 questions (10\%), and 8-12 questions (10\%). As shown in Figure \ref{fig:casenessgroups-wellbeing}, the impact of vaccination is monotonically increasing in individuals' prior levels of mental distress.

\subsubsection{Robustness} \label{sec:robustness}

Next we examine the robustness of our main results
to the inclusion of additional controls, the sample composition, alternative survey weights, and we explore the possibility of non-random attrition.

\paragraph{Additional controls}

In column (1) of Table \ref{table:robusttable}, we reproduce the results of our preferred specification, where we estimate equation (\ref{eq:did}) using invitations to instrument for vaccination.
In column (2), we allow for time-variant shocks affecting the 12 geographical regions of the UK by including in the specification a set of `age group*survey wave*region' fixed effects. The estimate is essentially the same (p.e.=12\% st. dev., s.e.=4\%), suggesting that regions where the roll-out was relatively faster were not exposed to unobserved time-variant shocks affecting psychological well-being. 

An additional potential threat to the validity of the analysis would be that the variation in the speed of the roll-out across different geographical areas is driven by underlying differences in incidence rates, which may affect psychological well-being. However, results are also unchanged when we control for the local incidence of COVID-19 at the time of survey (column 3), which is negatively correlated with psychological well-being but, as discussed above, is uncorrelated with vaccination rates.

\paragraph{Sample restrictions}
Our baseline sample includes individuals aged 40-80. In column 4 we restrict the sample to individuals aged 45-55 and 65-75, to minimize the possibility that the variation in invitation status that we exploit is driven by some unobserved factor which affects also psychological well-being. This sample restriction does not affect significantly our estimates (p.e.=13\% st. dev., s.e.=5\%).
Estimates are smaller, but significantly different from zero when we extend the analysis to all individuals in the survey, independently of their age (see column 5, p.e.=8\% st. dev., s.e.=3\%), when we only consider individuals who participated in all waves of the COVID survey (column 6, p.e.=7\% st. dev., s.e.=4\%) and when we give the same weight to all observations (column 7, p.e.=7\% st. dev., s.e.=2\%).

\paragraph{Non-random attrition}
A potential threat to the validity of the analysis is that vaccination may have affected participation in the survey. For instance, an upward bias would arise if individuals affected by vaccination side-effects were less likely to participate in the survey. To investigate this issue, we examine whether the probability of participation in the surveys conducted in January and March 2021 (waves 7 and 8) was lower for age groups with higher vaccination rates during the week leading up to the survey. 
We use administrative data from Public Health England, which provides information on vaccination rates by age group aggregated in fifteen 5-year intervals. To test for the existence of non-random attrition, we estimate the following equation using data from all survey waves:

\begin{equation} \label{eq:selection_bias}
\begin{split}
y_{at} &= \sum_{j \in \text{set of age groups}} \alpha_j \cdot I(j=a) + \sum_{w=2}^8 \beta_{w} \cdot I(w=wave_{at}) + \gamma \cdot VaccinationRate_{at} + \epsilon_{at} 
\end{split}
\end{equation}
where the outcome variable $y_{at}$ is the share of individuals in a given age group who participated in the survey conducted at $t$. The specification includes a set of age group fixed effects (1st term on the right-hand side) and survey wave fixed effects (2nd term). The main variable of interest, $VaccinationRate_{at}$, is the share of individuals in a given age group who were vaccinated during the week before the survey. We cluster standard errors at the level of age groups and, given the relatively low number of groups, we use bootstrapping.

If vaccination affects individuals' participation in the survey, we would expect this effect to be captured by coefficient $\gamma$. 
However, as shown in Table \ref{table:participation}, we do not observe any significant relationship between the share of individuals in a given age group who were vaccinated just before the survey wave and survey participation (p.e.=0.23, s.e.=0.14).

\subsubsection{Extensions}\label{sec:extensions}

In this section we provide a number of additional analyses. First, we examine the impact of vaccination across the different dimensions of the GHQ-12 index. Second, we study whether the impact is heterogeneous across different socio-economic groups. Third, we investigate whether there is any anticipation effect when individuals receive the invitation or make an appointment. Finally, we explore the impact of the second jab, using the information from a few hundred people who had received it by the time of the March 2021 survey.

\paragraph{Dimensions of the GHQ-12 index}

The GHQ-12 index aggregates information from 12 different questions. In Table \ref{table:dimensions}, we estimate the impact of vaccination separately for each question of the index. To ease the interpretation, similarly to what we did for the average GHQ-12 score, we standardize and invert each variable.
The largest impact of vaccination is observed on the ability to concentrate, the enjoyment of day-to-day activities, feeling of general happiness, lower likelihood of feeling unhappy or depressed, and of losing confidence in oneself. All these effects are statistically significant and the point estimate is above 0.10 st. dev.

\paragraph{Heterogeneity}

We examine the impact of vaccination separately for different groups of individuals according to gender, ethnicity, education, employment, family structure and number of friends. The effect of vaccination is slightly larger for women, and for individuals who are more educated, employed, not living with a partner and not working from home, but none of these differences is statistically significant at standard levels (see Figure \ref{fig:coef_het}).

\paragraph{Timing of the effect}

The survey information allows to identify four different phases of the vaccination process: (i) individuals who have not received an invitation yet for their first jab, (ii) individuals who have received an invitation but have not made an appointment, (iii) individuals who have the made the appointment but are not yet vaccinated, and (iv) individuals who have already received their first jab.

We explore the impact of each of these events -invitation, appointment and vaccination- using equation (\ref{eq:did}). As shown in Table \ref{table:appointment}, column 1, we do not observe any increase in the psychological well-being for individuals who received the invitation but did not make the appointment. However, we do observe a large increase in well-being for individuals who have received the invitation and made an appointment but are not yet vaccinated (p.e.=15\% st. dev., s.e.=7\%) and for individuals who have already received their first jab (p.e.=15\% st. dev., s.e.=7\%). The existence of a large positive effect already at the time of the appointment is consistent with an anticipation effect of the future consequences of vaccination.

\paragraph{Second vaccine}
So far we have focused on the impact of receiving the first jab. We investigate the impact of the second jab using the information provided by a few hundred individuals who had received the 2nd jab at the time of the March 2020 survey.
Among individuals older than 60, practically all of them (around 98\%) had received their first jab and 6\% had already received also their second jab.
As shown in column 7 of Table \ref{table:maintable}, the impact of the second jab is not significantly different from zero (p.e.= 2\% st. dev., s.e.=8\%), but the estimate is quite imprecise and we cannot reject that the effect was similar to the impact of the same jab.

\subsubsection{Regression discontinuity design}\label{sec:rdd}

\begin{figure}[htbp]
\caption{Regression discontinuity plots}
\begin{center}
\subfloat[Vaccination rate by age, March 2021.]{\includegraphics[height=2.3in]{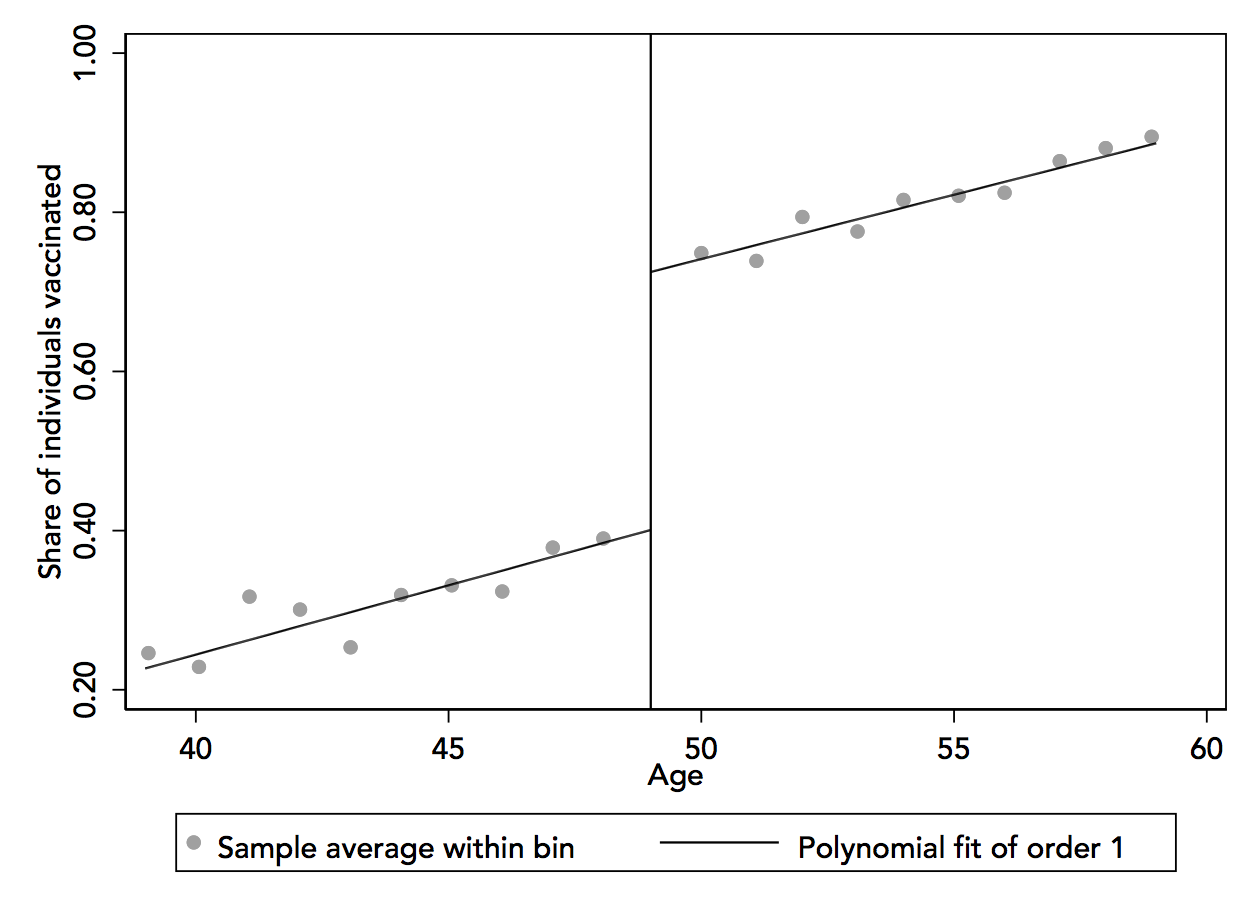}} \\
\subfloat[Mental distress by age, March 2021]{\includegraphics[height=2.3in]{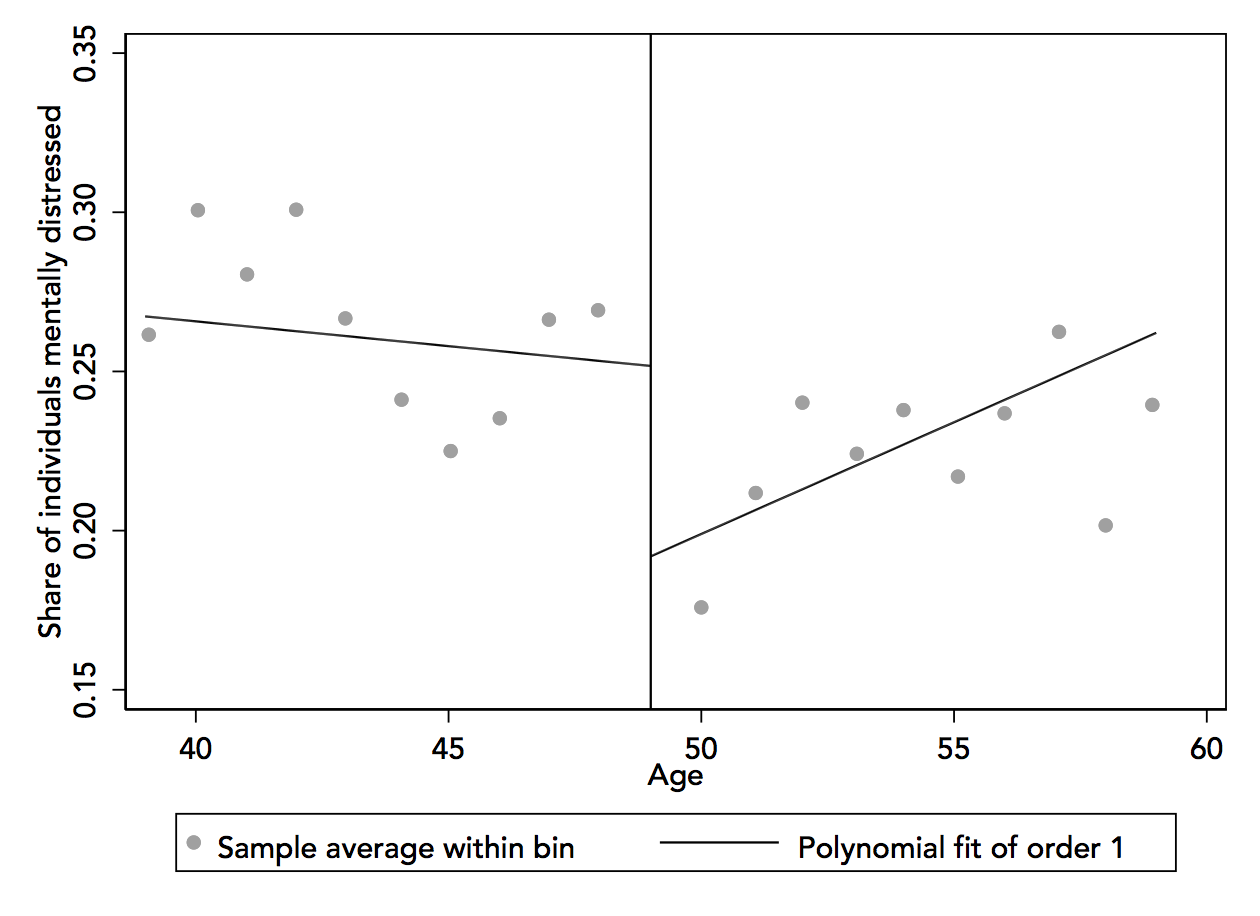}}
\subfloat[Placebo: Mental distress by age, April 2020-January 2021.]{\includegraphics[height=2.3in]{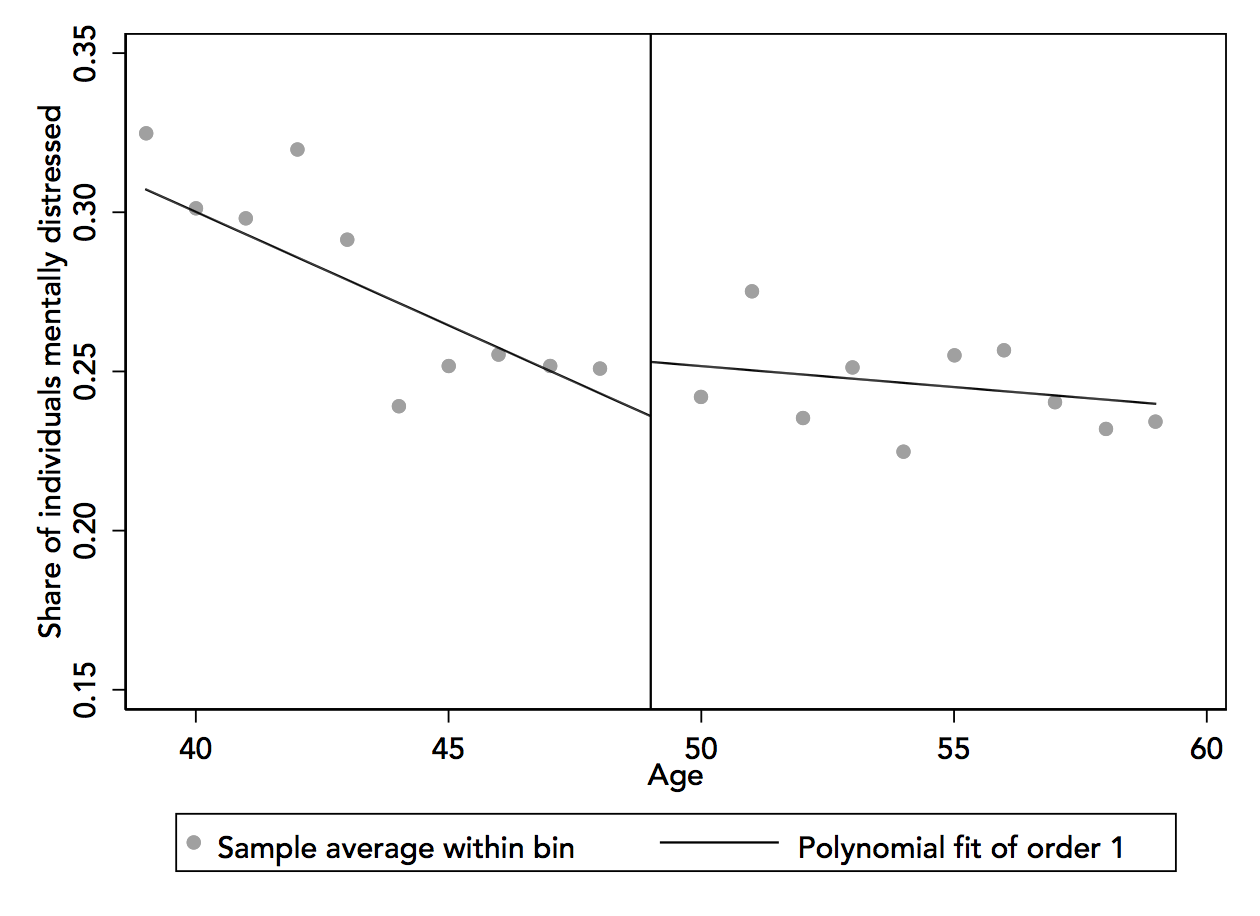}}
 \end{center}
 \vskip 2mm 
\textit{Notes:} In sub-figures (a) and (b) the sample includes all individuals aged 39-60 who participated in the 2021 March survey of \emph{Understanding Society}. Sub-figure (c) includes information from the seven survey waves that were conducted between April 2020 and January 2021. All three plots omit individuals aged 49 and in Health and Social Care occupations.
\label{fig:rdd}
\end{figure}

We also estimate the impact of vaccination exploiting the regression discontinuity design (RDD) observed in the March 2021 survey. In this wave, there exists a clear discontinuity in the age profile of people who have been vaccinated (see Figure \ref{fig:rdd}.a).\footnote{Instead, in the January 2021 survey vaccination rates increase linearly between age 65 and 75, without any visible discontinuity.} Just above age 49, around 90\% of individuals are vaccinated compared to only 45\% just below.
The existence of this discontinuity is likely to reflect that age 50 was the cut-off for the end of the first phase of the vaccination roll-out. We exploit this (fuzzy) RDD to estimate how vaccination affects psychological well-being. A potential advantage of using this RDD is that it relies on weaker assumptions than the DID-IV. Its consistency requires that there are no other relevant discontinuities at the 49 year old threshold, an assumption which in principle seems plausible and it is partially testable. On the flip side, a drawback of RDD is its more limited statistical power, as it exploits information only for individuals around the threshold.

We estimate the following equations using information from all respondents in the March 2021 survey wave:
\begin{align}
Vaccinated_{i} &= \beta_{firststage} \cdot I(age_i > 49) + f(age_i) + \epsilon_{i}  \tag{5.a}\label{eq:rdd-first}\\ 
y_{i} &= \beta_{reducedform} \cdot I(age_i > 49) + g(age_i) + \eta_{i} \tag{5.b}\label{eq:rdd-reduced} \\
y_{i} &= \beta_{fuzzy} \cdot \widehat{\text{Vaccinated}_i}+ h(age_i) + \rho_{i}  \tag{5.c}\label{eq:rdd-fuzzy}
\end{align}	

where $Vaccinated_{i}$ is a dummy variable capturing the vaccination status of individual $i$ in the March 2021 survey; $I(age_i > 49)$ is an indicator that takes value one if individual $i$ was more than 49 years old; $y_i$ is a measure of psychological well-being; $\widehat{Vaccinated}_i$ is the predicted value estimated in equation (\ref{eq:rdd-first}); and $f(age_i)$, $g(age_i)$ and $h(age_i)$ are flexible continuous functions that capture the relationship between the corresponding outcome variable and the \textit{running variable} age. To increase precision, we also include as a control the lagged value of the outcome variable.
We implement this regressions using the bandwidth selection procedure proposed by \cite{calonico2014robust}, local polynomials of order one at each side of the threshold, and a triangular kernel.\footnote{We exclude from the estimation individuals who are exactly 49, as their vaccination rate (66\%) is just in between the level of individuals aged 50-60 (around 94\%) and individuals aged 40-48 (around 45\%). We also exclude Health and Social workers, a group which was prioritized and had been vaccinated earlier.}

The optimal bandwidth is around 10 years in all estimations. The first stage estimation (equation \ref{eq:rdd-first}) shows that being above the 49 years of age threshold increases the probability of being vaccinated at the time of the March survey by 44 p.p. (see column 1 of Table \ref{table:rdd}) and, the estimation of the reduced form (equation \ref{eq:rdd-reduced}), that individuals above this threshold have 13\% st. dev. (s.e.=7\%) higher psychological well-being (column 2). According to the fuzzy RDD (equation \ref{eq:rdd-fuzzy}), vaccination increases psychological well-being by around 30\% st. dev. (s.e.=17\%). This is a substantial effect, twice as large as the DID-IV estimate, but it is imprecise and it is only marginally significant.
The estimate of the impact of vaccination on the probability of being mentally distressed is more precise. A discontinuity in the probability of being mentally distressed at the 49 years old threshold is clearly visible in the RD plot (Figure \ref{fig:rdd}.b). Individuals just above 49 years old are 10 p.p. (st.err.=4) less likely to be mentally distressed, implying that vaccination decreases the probability of being distressed by about 20 p.p. (s.e.=8) (columns 5 and 6).
A placebo analysis shows that the gap in psychological well-being observed at the 49 years threshold did not exist in previous waves, supporting that the difference observed in March 2021 reflects the impact of vaccination (see Figure \ref{fig:rdd}.c and Table \ref{table:rdd}, column 7). Furthermore, as expected, the age density function does not exhibit any discontinuity at the 49 years threshold (Figure \ref{fig:rdhistogram}).

In the lower panel of Table \ref{table:rdd} we conduct the same analysis restricting the sample to individuals who were mentally distressed in November of 2020. Consistent with our previous results, we find that the impact of vaccination is stronger in this sample. The probability of being distressed in March 2021 is 27 p.p. (s.e.=11) lower just above the 49 year threshold, and psychological well-being is 41\% st. dev. (s.e.=24\%) higher.

Overall, the fuzzy RDD estimates tend to be larger than the DID-IV ones. There are at least three potential explanations. First, note that the difference is not statistically significant. Second, each approach is identifying the impact on a different segment of the population. While in the DID-IV approach the estimate is obtained by comparing individuals of the same age and priority group who were invited or not for vaccination, in the fuzzy RDD approach we are comparing individuals just above and below the 49 years old threshold. The identity of `compliers' is also slightly different in both cases. The DID-IV identifies the impact of the vaccination for individuals who would get vaccinated if they receive an invitation. Instead, the fuzzy RDD provides information on the impact of the treatment for `compliers' around the threshold, i.e. individuals around 49 whose vaccination status varies depending on their age. 

\subsection{Persistence of the effect} \label{sec:mid-term}

The above evidence shows that vaccination has a large immediate effect on psychological well-being. Next we examine whether this effect persists at least during the period during which we can track respondents, between January and March 2021. More precisely, we compare the well-being of individuals vaccinated in January vs. the well-being of individuals of similar age who were vaccinated in February and March. If the effect of vaccines fades away over time, we would expect to observe a higher level of psychological well-being among individuals who have been vaccinated more recently.

We consider in our analysis all waves that took place in 2020 (before vaccination started) and the March 2021 wave (i.e. we exclude from this analysis the January 2020 wave), and we focus on the group of people who are more than 60 years old and less than 80, as for this group we can observe some variation in the timing of vaccination. Around one third of them had received their first jab in January 2021, and the rest got vaccinated during February and March. 
As shown in Table \ref{table:longterm}, the psychological well-being at the end of March 2021 of individuals who were vaccinated in January is slightly higher than the well-being of individuals who were vaccinated in February and March, although the difference is not statistically significant (p.e.=5\% st. dev., s.e.=4\%). This pattern suggests that the gains in psychological well-being do not fade away during the first two months, otherwise we would have observed a higher psychological well-being among individuals who have been vaccinated more recently.

\section{Discussion} \label{sec:discussion}

Using evidence from the UK vaccination roll-out, our analysis highlights that the benefits of vaccination are not limited to their impact on physical health. Vaccines have a large positive impact on psychological well-being, compensating for around one half of the decrease in psychological well-being caused by the pandemic. 
The effect is particularly large for mentally distressed individuals, who account for around one fourth of the adult population.

Our results strongly suggest that it might be convenient to prioritize the mentally distressed in vaccination roll-outs. 
An obvious advantage is that it would improve the overall psychological well-being of the population, a relevant dimension for policy makers. Moreover, mentally distressed people tend to have significantly higher utilisation rates of the health system. For instance, according to our data they are 50\% more likely to use outpatient and inpatient services.\footnote{Using information from wave 6 of the survey, before the vaccination roll-out, we observe that 31\% of mentally distressed individuals use NHS hospital and clinic outpatient services and 12\% in-patient services, compared to only 21\% and 8\% respectively in the rest of population.} Improving their psychological well-being might help to ease the pressure on the NHS.

There are several potential limitations of our study. 
First, we estimate only the direct impact of vaccinations on individuals who received the vaccination. Our estimate does not capture important indirect channels such as the positive externality provided by other individuals' vaccination.
Second, our analysis identifies the short-term of vaccination. In the longer term vaccinations might have additional positive effects, contributing for instance to the relaxation of the lock-downs that constraint labor market activity and social interactions.\footnote{Some authors have also pointed out that individuals typically over-report psychological well-being, and using the GHQ-12 score generally may lead to an underestimate of the effect of psychological distress on transitions into improved economic states \citep{brown2018mental}.}
More research is needed to determine the effects of vaccination on psychological well-being in the longer term.

\bibliographystyle{chicago}
\bibliography{bibliography}

\clearpage
\vfill

\pagebreak

\centerline{{\Large \textbf{Tables}}}

\overfullrule=0mm


\begin{table}[htbp]
\caption{Impact of vaccination on psychological well-being \label{table:maintable}} 
\scalebox{0.85}{
\centering
\begin{threeparttable} 
\begin{tabular}{l*{9}{c}} \hline\hline
                               &\multicolumn{1}{c}{(1)}   &\multicolumn{1}{c}{(2)}   &\multicolumn{1}{c}{(3)}   &\multicolumn{1}{c}{(4)}   &\multicolumn{1}{c}{(5)}   &\multicolumn{1}{c}{(6)}   &\multicolumn{1}{c}{(7)}   &\multicolumn{1}{c}{(8)}   \\
& OLS & DID & DID & DID & IV-DID & IV-DID & IV-DID & DID   \\ \addlinespace \hline \addlinespace 
1st vaccination                &   0.214***&   0.047** &   0.140***&   0.128***&   0.120***&   0.121***&   0.119*  &   0.108***\\
                               & (0.042)   & (0.024)   & (0.037)   & (0.033)   & (0.037)   & (0.036)   & (0.066)   & (0.032)   \\
2nd vaccination                &           &           &           &           &           &           &           &   0.019   \\
                               &           &           &           &           &           &           &           & (0.085)   \\
\addlinespace
First stage F                  &           &           &           &           &    5483   &    4314   &    1965   &           \\
N                              &  48,016   &  47,710   &  47,710   &  47,376   &  47,376   &  27,510   &  19,866   &  28,494   \\
\hline \addlinespace Wave FE & Yes & Yes & Yes & Yes & Yes & Yes & Yes & Yes  \\ Individual FE & No & Yes & Yes & Yes & Yes & Yes & Yes & Yes  \\ Wave*Age FE & No & No & Yes & Yes & Yes & Yes & Yes & Yes  \\ Wave*Priority FE & No & No & No & Yes & Yes & Yes & Yes & Yes \\ Sample & 40-80 & 40-80 & 40-80 & 40-80 & 40-80 & 61-80 & 40-60 & 61-80 \\ \hline \hline \end{tabular}

\begin{tablenotes} \footnotesize \item \textit{Notes:} The outcome variable is the standardized inverted GHQ-12 Likert score. All regressions include survey wave fixed effects. Columns 2-8 include individual fixed effects, columns 3-8 include a set of time event dummies interacted with age, and columns 4-8 a set of priority group dummies interacted with time event dummies and age groups.
In columns 5-7 vaccination is instrumented using invitations for vaccination. In column 6 we consider only the 61-80 age group and in column 7 the 40-60 age group. Column 8 considers both the 1st and 2nd vaccination for the 61-80 age group.
All regressions use sample weights and standard errors are clustered at the level of strata. 
* \(p<0.10\), ** \(p<0.05\), *** \(p<0.01\)
\end{tablenotes} 
\end{threeparttable}}
\end{table}


\begin{table}[htbp]
\caption{Mechanisms - COVID-19, health, and social life \label{table:mechanisms-health}}
\scalebox{0.90}{
\centering
\begin{threeparttable}
\setlength{\tabcolsep}{10pt}
\begin{tabular}{l*{8}{c}} \hline\hline
                    &\multicolumn{1}{c}{(1)}   &\multicolumn{1}{c}{(2)}   &\multicolumn{1}{c}{(3)}   &\multicolumn{1}{c}{(4)}   &\multicolumn{1}{c}{(5)}   &\multicolumn{1}{c}{(6)}   &\multicolumn{1}{c}{(7)}   \\
\multirow{ 2}{*}{Outcome:} & Had  & Tested  & General  & Risk  & \multirow{ 2}{*}{Loneliness} & Weekly & Daily  \\ & symptoms & positive & health & perception &  & walking & activities \\ \addlinespace \hline \addlinespace
1st vaccination     &       0.003   &       0.008   &       0.069*  &      -0.169***&      -0.081** &       0.149** &       0.152***\\
                    &     (0.011)   &     (0.008)   &     (0.037)   &     (0.059)   &     (0.038)   &     (0.063)   &     (0.058)   \\
\addlinespace
N                   &      47,363   &      47,376   &      15,202   &      43,029   &      47,365   &      13,848   &      47,376   \\
\hline \hline \end{tabular}

\begin{tablenotes} \footnotesize \item \textit{Notes:} In all columns we report estimates of equation (\ref{eq:did}), instrumenting the 1st vaccination with invitations, using sample weights and with standard errors clustered at the level of strata.
All regressions include individual fixed effects, age times time event fixed effects, and fixed effects for priority groups interacted with time event and age group dummies.
The outcome variables are: indicator for having had COVID-19 symptoms in column 1, having tested positive for COVID-19 in column 2, (standardized) self-assessed health in column 3 (based on question: "In general, would you say your health is...: excellent/very good/good/fair/poor"), (standardized) risk perception in column 4 (based on question: "In your view, how likely is it that you will contract COVID-19 in the next month? Very likely/Likely/Unlikely/Very unlikely"), (standardized) loneliness in column 5 (based on question: "how often one feels lonely: hardly ever or never/some of the time/often", and number of walking days in column 6 (based on question: "During the last 7 days, on how many days did you walk for at least 10 minutes at a time?"). In column 7 the outcome variable is an standardized measure of the answers to the question on the enjoyment of daily activities in the GHQ-12.
* \(p<0.10\), ** \(p<0.05\), *** \(p<0.01\)
\end{tablenotes}
\end{threeparttable}}
\end{table}


\clearpage
\vfill

\pagebreak

\centerline{{\Large \textbf{Figures}}}


\clearpage
\begin{appendices}

\section{Appendix: GHQ index}\label{sec:app-ghq}
The GHQ index is constructed as the sum of the following 12 different questions, each one scaled from 0 to 3 (0, not at all; 1, no more than usual; 2, rather more than usual; 3, much more than usual).

\begin{itemize}
\item \textbf{a. concentration}: Have you recently been able to concentrate on whatever you're doing?
\item \textbf{b. lack of sleep}: Have you recently lost much sleep over worry?
\item \textbf{c. playing a useful role}: Have you recently felt that you were playing a useful part in things?
\item \textbf{d. capable of making decisions}: Have you recently felt capable of making decisions about things?
\item \textbf{e. constantly under strain}: Have you recently felt constantly under strain?
\item \textbf{f. problem overcoming difficulties}: Have you recently felt you couldn't overcome your difficulties?
\item \textbf{g. enjoy day-to-day activities}: Have you recently been able to enjoy your normal day-to-day activities?
\item \textbf{h. ability to face problems}: Have you recently been able to face up to problems?
\item \textbf{i. unhappy or depressed}: Have you recently been feeling unhappy or depressed?
\item \textbf{j. losing confidence}: Have you recently been losing confidence in yourself?
\item \textbf{k. believe worthless}: Have you recently been thinking of yourself as a worthless person?
\item \textbf{l. general happiness}: Have you recently been feeling reasonably happy, all things considered?
\end{itemize}


\clearpage

\section{Appendix: Additional Tables}
\label{sec:additional_tables}
\setcounter{table}{0}
\renewcommand{\thetable}{A\arabic{table}}

\begin{table}[htbp]
\caption{Summary statistics \label{table:summary}}
\centering
\scalebox{0.75}{
\begin{threeparttable}
\begin{tabular}{l cccc|cc}
& \multicolumn{4}{c}{Population Above 16} & \multicolumn{2}{c}{Main sample} \\

            &           N&        Mean&         Min&         Max&           N&        Mean\\
\midrule
\emph{Individual characteristics}&            &            &            &            &            &            \\
Female      &      103819&        0.53&           0&           1&       48097&        0.49\\
Age         &      106227&       50.84&          16&         101&       48479&       60.64\\
White       &      106266&        0.90&           0&           1&       48479&        0.95\\
Born in UK  &      106266&        0.90&           0&           1&       48479&        0.92\\
Urban       &      105188&        0.75&           0&           1&       48479&        0.73\\
College     &      106266&        0.29&           0&           1&       48479&        0.27\\
Living with partner&      106266&        0.63&           0&           1&       48479&        0.72\\
Parent 0-15 children&       75231&        0.22&           0&           1&       36729&        0.18\\
\midrule
\emph{Priority groups}&            &            &            &            &            &            \\
Health or social worker&      106266&        0.12&           0&           1&       48479&        0.00\\
Shielded    &      106266&        0.10&           0&           1&       48479&        0.12\\
Cares for sick-disabled-elderly&      102064&        0.09&           0&           1&       47824&        0.09\\
Receiving formal care&       36229&        0.02&           0&           1&       16473&        0.02\\
Clinically vulnerable&      106266&        0.36&           0&           1&       48479&        0.43\\
Clinically extremely vulnerable&      106266&        0.05&           0&           1&       48479&        0.07\\
\midrule
\emph{Health and social life }&            &            &            &            &            &            \\
Psychological well-being GHQ-12&      106266&       12.64&           0&          36&       48479&       12.06\\
Mental distress&      106266&        0.26&           0&           1&       48479&        0.22\\
Had COVID-19 symptoms&      106220&        0.05&           0&           1&       48459&        0.04\\
Tested COVID-19 positive&      106266&        0.01&           0&           1&       48479&        0.01\\
General health&       35674&        2.64&           1&           5&       16192&        2.69\\
Loneliness  &      106226&        1.49&           1&           3&       48457&        1.40\\
Walking     &       35421&        5.04&           0&           7&       15651&        5.16\\
\midrule
\emph{Vaccine attitudes }&            &            &            &            &            &            \\
Unlikely to contract COVID19&       87391&        0.25&           0&           1&       43607&        0.28\\
Unlikely to get vaccinated&       87430&        0.17&           0&           1&       43650&        0.13\\
Not concerned with side effects&      106266&        0.92&           0&           1&       48479&        0.93\\
\midrule
\emph{Labour market }&            &            &            &            &            &            \\
Employed    &      106193&        0.60&           0&           1&       48449&        0.48\\
Weekly household income&       68938&      657.95&           0&        8000&       34828&      636.78\\
Hours worked&       61089&       27.62&           0&         100&       21163&       27.40\\
Works from home always or often&       61556&        0.36&           0&           1&       21274&        0.38\\
\midrule
\emph{Financial security }&            &            &            &            &            &            \\
Financial situation (current)&       68468&        2.04&           1&           5&       28419&        1.99\\
Savings amount&       34857&      248.66&           0&       20000&       15545&      233.42\\
Financial situation (future)&       68356&        2.02&           1&           3&       28376&        2.04\\
Marginal propensity to consume&       36782&        1.87&           1&           3&       16390&        1.86\\
\midrule
\emph{COVID-19 local incidence}&            &            &            &            &            &            \\
New cases per 100,000&      104565&       55.92&           0&        2533&       47817&       59.07\\

\hline \end{tabular}
 \begin{tablenotes} {\footnotesize \item \textit{Notes:} Columns 1-4 provide information for all survey respondents in waves 1-8 of the \emph{Understanding Society} COVID-19 survey. Columns 5-6 provides information for the main sample used in the paper, which includes individuals between 40 and 80 years old, and excludes Health and Social workers. The means are population-weighted. The information for priority groups is defined in wave 6 (November 2020).} 
\end{tablenotes}
\end{threeparttable}}
\end{table}

\begin{table}[htbp]
\caption{Variable definitions}
\label{table:variable_definitions}
\scalebox{0.70}{
\begin{threeparttable}
\setlength{\tabcolsep}{10pt}
\begin{tabular}{l|l|l}
 Variable name & Survey question & Scale \\ 
 \hline
 \addlinespace
 \addlinespace
 \textbf{cvinvite} - Invited for & Have you been invited to have the coronavirus vaccination & 1. Yes 2. No \\
 covid-19 vaccine & by the NHS (even if you have not had the vaccination yet)? & \\
 
 \addlinespace
 \addlinespace
 \textbf{hadcvvac} - Had covid-19 vaccine & Have you had a coronavirus vaccination? & 1. Yes, first vaccination only \\
 & & 2. Yes, both vaccinations \\
 & & 3. No, but I have an appointment 4. No \\

 \addlinespace
 \addlinespace
 \textbf{aidhh} - Cares for handicapped or & Is there anyone living with you who is sick, disabled or elderly & 1. Yes 2. No\\
 other in household & whom you look after or give special help to (for example, & \\
& a sick, disabled or elderly relative, husband, wife or friend
etc)? & \\

 \addlinespace
 \addlinespace
 \textbf{nhsshield} - NHS shielded patient & Have you received a letter, text or email from the NHS & 1. Yes 2. No \\
& or Chief Medical Officer saying that you have been identified & \\
& as someone at risk of severe illness if you catch coronavirus, & \\
& because you have an underlying disease or health condition? & \\
 
 \addlinespace
 \addlinespace
 \textbf{scsf1} - General health & In general, would you say your health is... & 1. Excellent 2. Very good 3. Good \\
 & & 4. Fair 5. Poor \\
 
 \addlinespace
 \addlinespace
 \textbf{hadsymp} - Has had symptoms & Have you experienced symptoms that could be caused & 1. Yes 2. No \\
 that could be coronavirus & by coronavirus (COVID-19) & \\
 
 \addlinespace
 \addlinespace
 \textbf{sclonely\_cv} - Loneliness & In the last 4 weeks, how often did you feel lonely? & 1. Hardly ever or never \\
 & & 2. Some of the time 3. Often \\
 
 \addlinespace
 \addlinespace
 \textbf{wday} - 7 days walking & During the last 7 days, on how many days did you walk & Numeric textbox: Days per week \\
& for at least 10 minutes at a time? & \\

 \addlinespace
 \addlinespace
 \textbf{wah} - Working at home & During the last four weeks how often did you work at home? & 1. Always 2. Often \\
 & & 3. Sometimes 4. Never \\
 
 \addlinespace
 \addlinespace
 \textbf{finnow} - Subjective financial & How well would you say you yourself are managing 
& 1. Living comfortably 2. Doing alright \\
 situation - current & financially these days? Would you say you are... & 3. Just about getting by \\ 
 & & 4. Finding it quite difficult \\ 
 
 \addlinespace
 \addlinespace
 \textbf{finfut\_cv3} - Subjective financial & Looking ahead, how do you think you will be 
& 1. Better off \\
 situation - future & financially 3 months from now, will you
be... & 2. Worse off than you are now \\ 
& & 3. Or about the same? \\

 \addlinespace
 \addlinespace
 \textbf{saved\_cv} - Savings amount & About how much have you personally managed & Numeric textbox: Pounds \\
 & to save in the last 4 weeks? & \\
 
 \addlinespace
 \addlinespace
 \textbf{mpc1} - Marginal propensity & Now consider a hypothetical situation where you unexpectedly & \\
to consume & receive a one-time payment of GBP500 today. & Over the next 3 months, I would: \\
 & We would like to know whether this extra income would cause & spend more than/the same as/less than \\
 & you to change your spending, borrowing and saving & than if I hadn’t received the GBP500 \\
 & behaviour in any way over the next 3 months. & \\
 
 \addlinespace
 \addlinespace
 \textbf{riskcv19} - Risk of getting covid19 & In your view, how likely is it that you will contract COVID-19 & 1. Very likely 2. Likely \\
 & in the next month? & 3. Unlikely 4. Very unlikely \\
 
 \addlinespace
 \addlinespace
 \textbf{vaxxer2} - Likelihood of taking up & When you are offered the coronavirus vaccination, & 1. Very likely 2. Likely \\
 a coronavirus vaccination & how likely or unlikely would you be to take it? & 3. Unlikely 4. Very unlikely \\
 
 \hline
\end{tabular}
\end{threeparttable}}
\end{table}


\thispagestyle{empty}
\begin{table}[htbp]
\centering
\caption{Predictors of Invitations and Vaccinations (1/2) \label{table:determinants}}
\scalebox{0.70}{
\begin{threeparttable}
\setlength{\tabcolsep}{15pt}
\begin{tabular}{l*{3}{c}} \hline\hline
                    &\multicolumn{1}{c}{(1)}         &\multicolumn{1}{c}{(2)}         \\
\multicolumn{1}{r}{\textbf{Outcome variable:}} & \multicolumn{1}{c}{\textbf{Invited}} & \multicolumn{1}{c}{\textbf{Vaccinated}} \\ \cline{2-3} \\ \hline \multicolumn{3}{l}{\emph{A: Priority groups}} \\ \addlinespace 
Shielded            &       0.097\sym{***}&       0.085\sym{***}\\
                    &     (0.029)         &     (0.021)         \\
Moderate risk       &       0.051\sym{***}&       0.066\sym{***}\\
                    &     (0.016)         &     (0.021)         \\
High risk           &       0.059\sym{**} &       0.076\sym{***}\\
                    &     (0.027)         &     (0.027)         \\
Cares for elderly/sick/disabled&       0.115\sym{***}&       0.119\sym{***}\\
                    &     (0.021)         &     (0.027)         \\

\hline

\addlinespace
\multicolumn{3}{l}{\emph{B: Individual characteristics}} \\ \addlinespace

Mental distress     &      -0.009         &      -0.007         \\
                    &     (0.015)         &     (0.018)         \\
Female              &       0.019         &      -0.001         \\
                    &     (0.012)         &     (0.014)         \\
White               &      -0.036         &      -0.022         \\
                    &     (0.034)         &     (0.034)         \\
Urban               &       0.012         &       0.001         \\
                    &     (0.015)         &     (0.018)         \\
College             &       0.011         &       0.026\sym{*}  \\
                    &     (0.013)         &     (0.015)         \\
Living with partner &       0.016         &       0.063\sym{***}\\
                    &     (0.015)         &     (0.018)         \\
Parent of children aged 0-15&      -0.007         &      -0.020         \\
                    &     (0.024)         &     (0.026)         \\

\hline

\multicolumn{3}{l}{\emph{C: Vaccine attitudes and risk perception}} \\ \addlinespace

Risk attitude       &      -0.000         &       0.001         &            \\
                    &     (0.006)         &     (0.007)         &            \\
Vaccine sceptic in Nov 2020&      -0.020         &      -0.024         &            \\
                    &     (0.016)         &     (0.016)         &            \\
Not concerned with vaccine side effects&      -0.028         &       0.080\sym{**} &            \\
                    &     (0.029)         &     (0.032)         &            \\

\hline

\multicolumn{3}{l}{\emph{D: COVID-19 incidence and regions}} \\ \addlinespace

COVID-19 cases      &      -0.003         &       0.004         &            \\
                    &     (0.006)         &     (0.006)         &            \\
North West          &       0.066\sym{**} &       0.017         &            \\
                    &     (0.029)         &     (0.033)         &            \\
Yorkshire and The Humber&       0.036         &      -0.020         &            \\
                    &     (0.026)         &     (0.036)         &            \\
East Midlands       &       0.042         &      -0.006         &            \\
                    &     (0.029)         &     (0.038)         &            \\
West Midlands       &       0.047\sym{*}  &      -0.043         &            \\
                    &     (0.028)         &     (0.038)         &            \\
East of England     &       0.074\sym{**} &      -0.014         &            \\
                    &     (0.030)         &     (0.032)         &            \\
London              &       0.114\sym{***}&      -0.002         &            \\
                    &     (0.030)         &     (0.036)         &            \\
South East          &       0.036         &       0.016         &            \\
                    &     (0.024)         &     (0.033)         &            \\
South West          &       0.014         &      -0.019         &            \\
                    &     (0.028)         &     (0.036)         &            \\
Wales               &      -0.098\sym{***}&      -0.113\sym{**} &            \\
                    &     (0.035)         &     (0.046)         &            \\
Scotland            &      -0.066\sym{*}  &      -0.110\sym{***}&            \\
                    &     (0.035)         &     (0.041)         &            \\
Northern Ireland    &      -0.034         &      -0.060         &            \\
                    &     (0.035)         &     (0.047)         &            \\

\hline \end{tabular}

\end{threeparttable}}
\end{table}

\begin{table}[htbp]
\centering
\caption*{Table A3: Predictors of Invitations and Vaccinations (2/2)}
\scalebox{0.70}{
 \begin{threeparttable}
\setlength{\tabcolsep}{15pt}
\begin{tabular}{l*{3}{c}} \hline
\multicolumn{3}{l}{\emph{E: Key worker sectors}} \\ \addlinespace
Health and social care&       0.235\sym{***}&       0.204\sym{***}\\
                    &     (0.047)         &     (0.054)         \\
Education and childcare&       0.019         &       0.076         \\
                    &     (0.060)         &     (0.073)         \\
Key public services &       0.011         &       0.013         \\
                    &     (0.052)         &     (0.062)         \\
Local and national government&      -0.048         &      -0.008         \\
                    &     (0.056)         &     (0.066)         \\
Food and other necessary goods&      -0.017         &       0.019         \\
                    &     (0.046)         &     (0.058)         \\
Public safety and national security&      -0.061         &      -0.030         \\
                    &     (0.061)         &     (0.077)         \\
Transport           &      -0.010         &       0.048         \\
                    &     (0.059)         &     (0.071)         \\
Utilities, communications and financial&      -0.040         &       0.009         \\
                    &     (0.055)         &     (0.065)         \\

\hline

\multicolumn{3}{l}{\emph{F: Occupation sectors}} \\ \addlinespace

Agriculture, forestry, fishing&       0.029         &      -0.000         \\
                    &     (0.070)         &     (0.084)         \\
Mining and quarrying&       0.267         &       0.253         \\
                    &     (0.173)         &     (0.193)         \\
Manufacturing       &      -0.055         &      -0.140\sym{**} \\
                    &     (0.045)         &     (0.063)         \\
Electricity, gas, steam&      -0.046         &      -0.249\sym{**} \\
                    &     (0.064)         &     (0.118)         \\
Water supply, sewage, waste&      -0.061         &      -0.122         \\
                    &     (0.097)         &     (0.113)         \\
Construction        &      -0.037         &      -0.134\sym{**} \\
                    &     (0.049)         &     (0.055)         \\
Wholesale and retail trade&       0.007         &      -0.042         \\
                    &     (0.044)         &     (0.053)         \\
Repair of motor vehicles,  motocycles&      -0.160\sym{*}  &      -0.225\sym{**} \\
                    &     (0.085)         &     (0.094)         \\
Transportation and storage&      -0.023         &      -0.131\sym{**} \\
                    &     (0.059)         &     (0.065)         \\
Accommodation and food service&      -0.022         &      -0.061         \\
                    &     (0.062)         &     (0.076)         \\
Information and communication&      -0.079         &      -0.137\sym{**} \\
                    &     (0.055)         &     (0.060)         \\
Financial and insurance activities&      -0.021         &      -0.066         \\
                    &     (0.051)         &     (0.060)         \\
Real estate activities&      -0.012         &      -0.032         \\
                    &     (0.054)         &     (0.064)         \\
Professional, scientific, technical&      -0.025         &      -0.065         \\
                    &     (0.046)         &     (0.054)         \\
Administrative and support service&       0.090\sym{*}  &       0.041         \\
                    &     (0.054)         &     (0.060)         \\
Public administration and defence&       0.021         &      -0.039         \\
                    &     (0.050)         &     (0.061)         \\
Education           &       0.013         &      -0.058         \\
                    &     (0.049)         &     (0.061)         \\
Human health and social work&       0.095\sym{*}  &       0.017         \\
                    &     (0.049)         &     (0.055)         \\
Arts, entertainment, recreation&      -0.020         &      -0.071         \\
                    &     (0.069)         &     (0.078)         \\
Other service activities&       0.022         &       0.003         \\
                    &     (0.046)         &     (0.053)         \\
Activities of households as employers&       0.059         &       0.058         \\
                    &     (0.100)         &     (0.102)         \\
\addlinespace
Observations        &        6835         &        4556         \\
\hline\hline \end{tabular}

\begin{tablenotes} \item \textit{Notes:} We report OLS estimates using sample weights and standard errors clustered at the level of strata. In column 1 the sample includes respondents in the January and March 2021 survey waves of `Understanding Society' and, in column 2, we individuals in these waves who had received an invitation for vaccination. The outcome variable is an indicator for being invited for vaccination in column 1 and an indicator for being vaccinated or having an appointment in column 2. All specifications also include fixed effects for age and survey date. Omitted category: male, non-white, from rural area, without a degree, not living with a partner, not a parent of children aged 0-15 with not applicable industry category from the North East, not a key worker, not clinically vulnerable, not shielding, not caring for vulnerable others, not receiving care and extremely likely to take up the vaccine and unlikely / very unlikely to get COVID19. * (p<0.10), ** (p<0.05), *** (p<0.01)
\end{tablenotes}
 \end{threeparttable}
 }
\end{table}


\begin{table}[htbp]
\caption{Impact of vaccination on mental distress \label{table:maintable-distress}} 
\scalebox{0.85}{
\centering
\begin{threeparttable} 
\begin{tabular}{l*{9}{c}} \hline\hline
                               &\multicolumn{1}{c}{(1)}   &\multicolumn{1}{c}{(2)}   &\multicolumn{1}{c}{(3)}   &\multicolumn{1}{c}{(4)}   &\multicolumn{1}{c}{(5)}   &\multicolumn{1}{c}{(6)}   &\multicolumn{1}{c}{(7)}   &\multicolumn{1}{c}{(8)}   \\
& OLS & DID & DID & DID & IV-DID & IV-DID & IV-DID & DID  \\ \addlinespace \hline \addlinespace 
1st vaccination                &  -0.056***&  -0.012   &  -0.039** &  -0.036*  &  -0.043** &  -0.035   &  -0.051   &  -0.016   \\
                               & (0.017)   & (0.014)   & (0.019)   & (0.019)   & (0.021)   & (0.027)   & (0.033)   & (0.025)   \\
2nd vaccination                &           &           &           &           &           &           &           &   0.040   \\
                               &           &           &           &           &           &           &           & (0.077)   \\
\addlinespace
First stage F                  &           &           &           &           &    5483   &    4314   &    1965   &           \\
N                              &  48,016   &  47,710   &  47,710   &  47,376   &  47,376   &  27,510   &  19,866   &  28,494   \\
\hline \addlinespace Wave FE & Yes & Yes & Yes & Yes & Yes & Yes & Yes & Yes  \\ Individual FE & No & Yes & Yes & Yes & Yes & Yes & Yes & Yes  \\ Wave*Age FE & No & No & Yes & Yes & Yes & Yes & Yes & Yes  \\ Wave*Priority FE & No & No & No & Yes & Yes & Yes & Yes & Yes  \\ Sample & 40-80 & 40-80 & 40-80 & 40-80 & 40-80 & 61-80 & 40-60 & 61-80  \\ \hline \hline \end{tabular}

\begin{tablenotes} \footnotesize \item \textit{Notes:} The outcome variable is an indicator for having clinically significant levels of mental distress. All regressions include survey wave fixed effects. Columns 2-8 include individual fixed effects, columns 3-8 include a set of time event dummies interacted with age, and columns 4-8 a set of priority group dummies interacted with time event dummies and age groups.
In columns 5-7 vaccination is instrumented using invitations for vaccination. In column 6 we consider only the 61-80 age group and in column 7 the 40-60 age group. Column 8 considers both the 1st and 2nd vaccination for the 61-80 age group.
All regressions use sample weights and standard errors are clustered at the level of strata.  * \(p<0.10\), ** \(p<0.05\), *** \(p<0.01\)
\end{tablenotes} 
\end{threeparttable}}
\end{table}


\begin{table}[htbp]
\caption{Labor market and household finances \label{table:labour-financial}}
\scalebox{0.80}{
\centering
\begin{threeparttable}
\setlength{\tabcolsep}{10pt}
\begin{tabular}{l*{9}{c}} \hline\hline
               &\multicolumn{1}{c}{(1)}   &\multicolumn{1}{c}{(2)}   &\multicolumn{1}{c}{(3)}   &\multicolumn{1}{c}{(4)}   &\multicolumn{1}{c}{(5)}   &\multicolumn{1}{c}{(6)}   &\multicolumn{1}{c}{(7)}   &\multicolumn{1}{c}{(8)}   \\
Outcomes: & Employed  & Hours & Weekly & Home & Financial & Savings & Financial & Marginal \\ &  & worked & income & working & situation & amount & situation & propensity \\ & & & & & (current) & & (future) & to consume \\ \addlinespace \hline \addlinespace
1st vaccination&    0.006   &   -2.434*  &   -0.231   &   -0.007   &    0.098*  &   -0.041   &   -0.057   &    0.066   \\
               &  (0.010)   &  (1.424)   &  (0.186)   &  (0.021)   &  (0.056)   &  (0.091)   &  (0.109)   &  (0.097)   \\
\addlinespace
N              &   19,859   &   15,581   &   15,302   &   15,653   &   27,337   &   14,280   &   27,311   &   15,555   \\
\hline \hline \end{tabular}

\begin{tablenotes} \footnotesize \item \textit{Notes:} In all columns we report estimates of equation (\ref{eq:did}), instrumenting the 1st vaccination with invitation, using sample weights and with standard errors clustered at the level of strata. All regressions include individual fixed effects, age times time event fixed effects, and fixed effects for priority groups interacted with time event and age group dummies.
The outcome variables are: an indicator for being employed in column 1, hours worked in column 2, log weekly income in column 3 and an indicator for working from home in column 4, subjective financial situation now in column 5: "How well would you say you yourself are managing financially these days? Would you say you are...", amount saved in column 6, subjective financial situation in the future in column 7, and marginal propensity to consume in column 8.
* \(p<0.10\), ** \(p<0.05\), *** \(p<0.01\)
\end{tablenotes}
\end{threeparttable}}
\end{table}


\begin{table}[htbp]
\centering
\caption{Mechanisms - Vaccine attitudes \label{table:mech-vax}}
\scalebox{0.80}{
\begin{threeparttable}
\setlength{\tabcolsep}{15pt}
\begin{tabular}{l*{4}{c}} \hline\hline
                                        &\multicolumn{1}{c}{(1)}   &\multicolumn{1}{c}{(2)}   &\multicolumn{1}{c}{(3)}   \\
& First stage & Reduced form & IV \\ \hline \addlinespace Outcome: & 1st vaccination & Well-being & Well-being \\ \addlinespace \hline \addlinespace
Invited                                 &    0.745***&   -0.094   &            \\
                                        &  (0.074)   &  (0.102)   &            \\
No side effects * Invited               &    0.174** &    0.224** &            \\
                                        &  (0.073)   &  (0.110)   &            \\
1st vaccination                         &            &            &   -0.126   \\
                                        &            &            &  (0.138)   \\
No side effects * 1st vaccination       &            &            &    0.268*  \\
                                        &            &            &  (0.145)   \\
\addlinespace
N                                       &   47,700   &   47,700   &   47,700   \\
\hline \hline \end{tabular}

\begin{tablenotes} \footnotesize \item \textit{Notes:} All regressions use sample weights and standard errors are clustered at the level of strata. The variable `No side effects' is an indicator for individuals who did not report any concerns with vaccines' potential side effects in the November 2020 survey wave. * \(p<0.10\), ** \(p<0.05\), *** \(p<0.01\)
\end{tablenotes}
\end{threeparttable}}
\end{table}

\begin{table}[htbp]
\caption{Impact on individuals with clinically significant levels of mental distress \label{table:mentally-distressed}}
\scalebox{0.78}{
\centering
\begin{threeparttable}
\setlength{\tabcolsep}{15pt}
\begin{tabular}{l*{8}{c}} \hline\hline
                    &\multicolumn{1}{c}{(1)}   &\multicolumn{1}{c}{(2)}   &\multicolumn{1}{c}{(3)}   &\multicolumn{1}{c}{(4)}   &\multicolumn{1}{c}{(5)}   &\multicolumn{1}{c}{(6)}   \\
Outcome: & Well-being & Well-being & Distress & Perceived risk & Loneliness & Daily activities \\ \addlinespace \hline \addlinespace
1st vaccination     &       0.138** &       0.119*  &      -0.034   &      -0.261***&      -0.109** &       0.113*  \\
                    &     (0.061)   &     (0.061)   &     (0.028)   &     (0.060)   &     (0.043)   &     (0.068)   \\
Mentally distressed &               &       0.173***&      -0.095***&      -0.057   &      -0.097** &       0.149*  \\
                    &               &     (0.066)   &     (0.035)   &     (0.064)   &     (0.048)   &     (0.079)   \\
\addlinespace
N                   &      15,205   &      14,238   &      14,238   &      14,146   &      14,232   &      14,238   \\
\hline \hline \end{tabular}

\begin{tablenotes} \footnotesize \item \textit{Notes:} The  table reports the results of estimating the IV equation for the sample of individuals who were mentally distressed in September 2020, using data for the survey waves in November 2020, January 2021 and March 2021. All regressions include using sampling weights, individual fixed effects, age times time event fixed effects, and fixed effects for priority groups interacted with time event and age group dummies. Standard errors are clustered at the level of strata. The outcome variable is the standardized inverted GHQ-12 Likert score in columns 1-2, an indicator for being mentally distressed in column 3, the (standardized) risk perception in column 4 (based on question: "In your view, how likely is it that you will contract COVID-19 in the next month? Very likely/Likely/Unlikely/Very unlikely"), (standardized) loneliness in column 5 (based on question: "how often one feels lonely: hardly ever or never/some of the time/often" and enjoying daily activities in column 6, which is one of the GHQ-12 dimensions. * \(p<0.10\), ** \(p<0.05\), *** \(p<0.01\)
\end{tablenotes}
\end{threeparttable}}
\end{table}

\begin{table}[htbp]
\caption{Robustness analysis \label{table:robusttable}}
\scalebox{0.85}{
\centering
\begin{threeparttable}
\setlength{\tabcolsep}{10pt}
\begin{tabular}{l*{7}{c}} \hline\hline
               &\multicolumn{1}{c}{(1)}   &\multicolumn{1}{c}{(2)}   &\multicolumn{1}{c}{(3)}   &\multicolumn{1}{c}{(4)}   &\multicolumn{1}{c}{(5)}   &\multicolumn{1}{c}{(6)}   &\multicolumn{1}{c}{(7)}   \\
& Main & Regional shocks & Local cases &  5-years & All sample & Balanced & No weights \\ \addlinespace \hline \addlinespace
1st vaccination&    0.120***&    0.116***&    0.129***&    0.128***&    0.085***&    0.068*  &    0.070***\\
               &  (0.037)   &  (0.036)   &  (0.038)   &  (0.046)   &  (0.028)   &  (0.036)   &  (0.022)   \\
COVID-19 cases &            &            &   -0.014** &            &            &            &            \\
               &            &            &  (0.006)   &            &            &            &            \\
\addlinespace
N              &   47,376   &   47,374   &   47,172   &   27,676   &   66,560   &   36,530   &   54,899   \\
\hline \addlinespace Sample & 40-80 & 40-80 & 40-80 & 45-55, 65-75 & all & 40-80 & 40-80  \\ \hline \hline \end{tabular}

\begin{tablenotes} \footnotesize \item \textit{Notes:} The table reports the results of estimating equation (\ref{eq:did}) using sampling weights and instrumenting vaccination with invitation. The outcome variable is the standardized inverted GHQ-12 Likert score. All regressions include individual fixed effects, age times time event fixed effects, and fixed effects for priority groups interacted with time event and age group dummies. Column 3 allows for time variant shocks at the regional level. Column 4 introduces the local COVID-19 incidence rate. Column 5 uses a narrower age bandwidth: 45-55 and 65-75, respectively. Column 6 removes any age sample restrictions. Column 7 considers the longitudinal weights for a balanced panel estimation and in column 7 there are no weights. Standard errors clustered at the level of strata. * \(p<0.10\), ** \(p<0.05\), *** \(p<0.01\) \end{tablenotes}
\end{threeparttable}}
\end{table}

\begin{table}[htbp]
\caption{Attrition \label{table:participation}}
\centering
\begin{threeparttable}
\setlength{\tabcolsep}{18pt}
\begin{tabular}{l*{1}{c}} \hline\hline \\ Outcome: & Participation \\
\addlinespace \hline \addlinespace
Vaccination rate         &    0.226   \\
                         &  (0.138)   \\
\addlinespace
N                        &      120   \\
\hline \addlinespace \hline \hline \end{tabular}

\begin{tablenotes} \footnotesize \item \textit{Notes:} Population sample in the COVID-19 survey of Understanding Society, collapsed by wave and age groups in 5-year intervals to match with administrative vaccination uptake data. The outcome variable is the share of people in the age group who have participated in a given survey wave. The vaccination rate is the percentage of people in that age group who have been vaccinated in the week leading up to the survey. The regression includes wave and age groups fixed effects. Bootstrapped standard errors clustered at the level of age groups. * \(p<0.10\), ** \(p<0.05\), *** \(p<0.01\)
\end{tablenotes}
\end{threeparttable}
\end{table}


\begin{table}[htbp]
\caption{Dimensions of psychological well-being \label{table:dimensions}}
\scalebox{0.90}{
\centering
\begin{threeparttable} 
\setlength{\tabcolsep}{10pt}
\begin{tabular}{l*{6}{c}} \hline\hline
               &\multicolumn{1}{c}{(a)}&\multicolumn{1}{c}{(b)}&\multicolumn{1}{c}{(c)}&\multicolumn{1}{c}{(d)}&\multicolumn{1}{c}{(e)}&\multicolumn{1}{c}{(f)}\\
& Concentration & Lack of & Playing a & Capable of & Constantly & Overcome  \\ &  & sleep & useful role & decisions & under strain & difficulties \\ \addlinespace \hline \addlinespace
1st vaccination&    0.135** &    0.023   &    0.059   &    0.092   &    0.087** &    0.038   \\
               &  (0.054)   &  (0.042)   &  (0.045)   &  (0.063)   &  (0.040)   &  (0.043)   \\
\addlinespace
N              &   47,376   &   47,376   &   47,376   &   47,376   &   47,376   &   47,376   \\

 \hline
 \addlinespace

&(g)&(h)&(i)&(j)&(k)&(l) \\ & Enjoy daily & Ability to & Unhappy or & Losing & Believe & General  \\ & activities & face problems & depressed & confidence & worthless & happiness \\ \addlinespace \hline \addlinespace
1st vaccination&    0.152***&    0.062   &    0.126***&    0.133***&    0.015   &    0.135***\\
               &  (0.058)   &  (0.058)   &  (0.048)   &  (0.046)   &  (0.040)   &  (0.046)   \\
\addlinespace
N              &   47,376   &   47,376   &   47,376   &   47,376   &   47,376   &   47,376   \\
\hline \hline \end{tabular}

\begin{tablenotes} \footnotesize \item \textit{Notes:} The table reports the results of estimating equation (\ref{eq:did}) using sampling weights and instrumenting vaccination with invitation. In each cell the outcome variable is a standardized single dimension of the GHQ-12 Likert index. All regressions include individual fixed effects, age times time event fixed effects, and fixed effects for priority groups interacted with time event and age group dummies. Standard errors clustered at the level of strata. * \(p<0.10\), ** \(p<0.05\), *** \(p<0.01\)
\end{tablenotes}
\end{threeparttable}}
\end{table}

\begin{table}[htbp]
\caption{Timing of the effect: invitation, appointment and vaccination \label{table:appointment}}
\centering
\scalebox{0.90}{
\begin{threeparttable} 
\setlength{\tabcolsep}{10pt}
\begin{tabular}{l*{3}{c}} \hline\hline
                               &\multicolumn{1}{c}{(1)}   &\multicolumn{1}{c}{(2)}   \\
\multirow{2}{*}{Outcome var.:} & Psychological  & Mentally \\ & well-being & distressed \\ \addlinespace \hline \addlinespace
Vaccinated                     &   0.152** &  -0.018   \\
                               & (0.068)   & (0.036)   \\
Appointment                    &   0.148** &  -0.021   \\
                               & (0.070)   & (0.041)   \\
Invited                        &  -0.028   &  -0.022   \\
                               & (0.068)   & (0.036)   \\
\addlinespace
N                              &  47,376   &  47,376   \\
 \\ \hline \hline \end{tabular}

\begin{tablenotes} \footnotesize \item \textit{Notes:} The table reports the results of estimating equation (\ref{eq:did}) using sampling weights and instrumenting vaccination with invitation. In column 1 the outcome variable is the standardized inverted GHQ-12 Likert score, and in column 2 an indicator for being mentally distressed. All regressions include individual fixed effects, age times time event fixed effects, and fixed effects for priority groups interacted with time event and age group dummies. Standard errors clustered at the level of strata. * \(p<0.10\), ** \(p<0.05\), *** \(p<0.01\)
\end{tablenotes}
\end{threeparttable}}
\end{table}


\begin{table}[htbp]
\caption{Regression discontinuity design\label{table:rdd}}
\scalebox{0.90}{
\centering
\begin{threeparttable}
\setlength{\tabcolsep}{8pt}
\begin{tabular}{l*{7}{c}} \hline\hline
               &\multicolumn{1}{c}{(1)}   &\multicolumn{1}{c}{(2)}   &\multicolumn{1}{c}{(3)}   &\multicolumn{1}{c}{(4)}   &\multicolumn{1}{c}{(5)}   &\multicolumn{1}{c}{(6)}   &\multicolumn{1}{c}{(7)}   \\
& First stage & RDD & Fuzzy RDD & Placebo & RDD & Fuzzy RDD & Placebo \\ \hline \addlinespace Survey Wave: & March 2021 & March 2021 & March 2021 & Waves 1-7 & March 2021 & March 2021 & Waves 1-7 \\ \hline \addlinespace Outcome: & 1st vaccination & Well-being & Well-being & Well-being & Distress & Distress & Distress \\ \hline    \multicolumn{7}{l}{\emph{Panel A: All}} \\ \addlinespace 
Age>49         &    0.445***&    0.132*  &    0.296*  &   -0.017   &   -0.101***&   -0.198** &    0.014   \\
               &  (0.034)   &  (0.073)   &  (0.166)   &  (0.024)   &  (0.037)   &  (0.077)   &  (0.011)   \\
\addlinespace
Bandwidth      &      9.7   &      8.6   &      8.9   &     10.2   &      8.5   &     10.0   &     11.6   \\
N              &   10,388   &    8,864   &    8,864   &   60,098   &    8,864   &    8,864   &   60,098   \\
\hline

\multicolumn{7}{l}{\emph{Panel B: Mentally distressed}} \\ \addlinespace

Age>49         &    0.435***&    0.406*  &    0.768   &    0.086   &   -0.271** &   -0.435*  &   -0.046   \\
               &  (0.063)   &  (0.238)   &  (0.575)   &  (0.087)   &  (0.113)   &  (0.255)   &  (0.034)   \\
\addlinespace
Bandwidth      &     12.8   &      7.1   &      8.9   &      7.6   &      6.5   &      9.3   &      9.1   \\
N              &    2,097   &    1,913   &    1,913   &   12,513   &    1,913   &    1,913   &   12,513   \\
\hline \hline \end{tabular}

\begin{tablenotes} \footnotesize \item \textit{Notes:} The outcome variable is an indicator for being vaccinated in column 1, the standardized inverted GHQ Likert score in columns 2-4, and an indicator for being mentally distress at the time of the survey in columns 5-7. Column 1 estimates the first stage for the probability of being vaccinated above the 49 years old age threshold. Columns 2 and 5 present reduced form estimations, and columns 3 and 6 show the fuzzy RDD, where vaccination is instrumented using the age threshold. Columns 4 and 7 implement placebo reduced form regressions using information from waves 1-7. Standard errors clustered at the level of strata. * \(p<0.10\), ** \(p<0.05\), *** \(p<0.01\)
\end{tablenotes}
\end{threeparttable}}
\end{table}

\begin{table}[htbp]
\caption{Persistence of the effect \label{table:longterm}}
\centering
\begin{threeparttable}
\setlength{\tabcolsep}{18pt}
\begin{tabular}{l*{1}{c}} \hline\hline
                                   &\multicolumn{1}{c}{(1)}   \\
\addlinespace \hline \addlinespace
Vaccinated in January 2021         &    0.053   \\
                                   &  (0.046)   \\
\addlinespace
N                                  &   28,948   \\
\hline \addlinespace Sample: & 61-80 \\ \hline \hline \end{tabular}

\begin{tablenotes} \footnotesize \item \textit{Notes:} The sample includes information from individuals aged 61-80 in waves 1-6 and 8.
The outcome variable is the standardized inverted GHQ-12 Likert score. The variable `Vaccinated in January 2021' takes value 1 for individuals who received their first vaccination before the January 2021 survey wave and zero otherwise. Individuals in the latter group were mostly vaccinated in February and March 2021. Standard errors clustered at the level of strata. * \(p<0.10\), ** \(p<0.05\), *** \(p<0.01\)
\end{tablenotes}
\end{threeparttable}
\end{table}

\clearpage

\section{Appendix: Additional Figures}
\label{sec:additional_figures}
\setcounter{figure}{0}
\renewcommand{\thefigure}{A\arabic{figure}}

\begin{figure}[htbp]
\caption{COVID-19 vaccination roll-out plan of Public Health England}
\begin{center}
 \includegraphics[width=.58\linewidth]{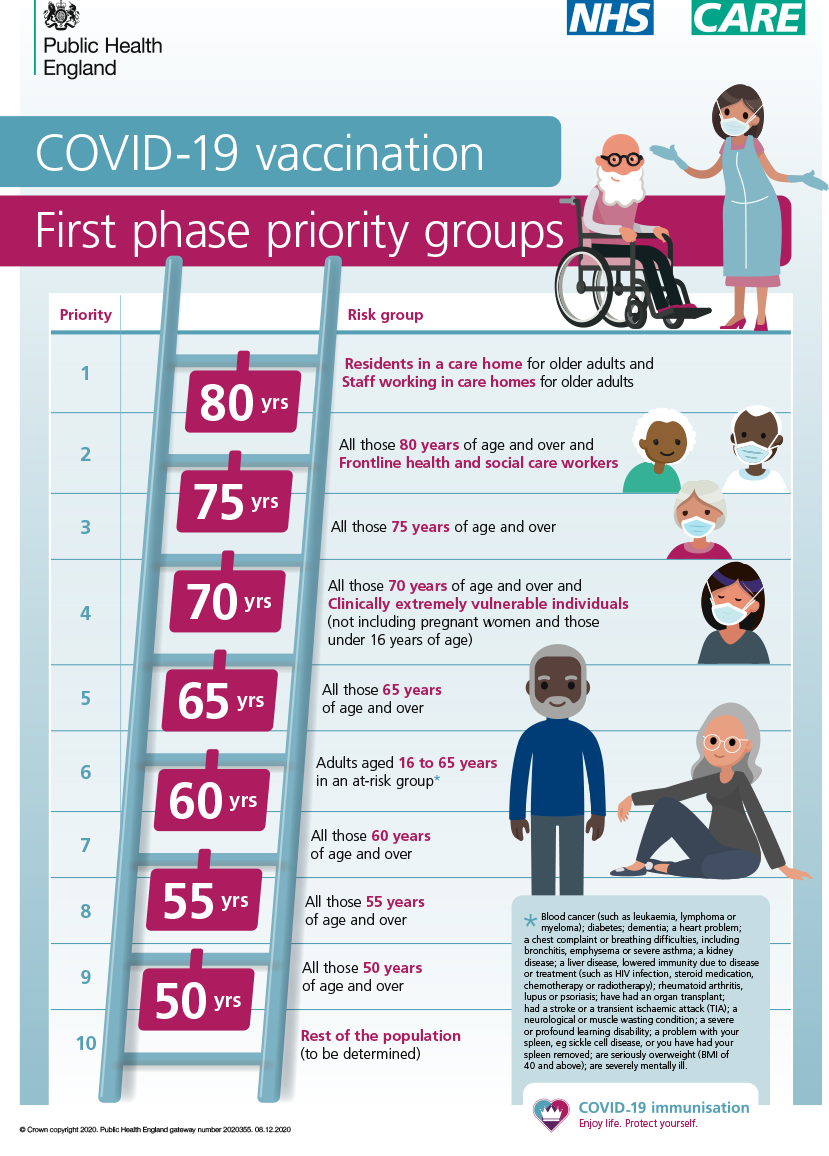}
 \end{center}
\caption*{\textit{Notes:} The priorities for the COVID-19 vaccination programme were established based on the independent report of the Joint Committee on Vaccination and Immunisation (JCVI) of December 30, 2020.}
\label{fig:priority-poster}
\end{figure}

\begin{figure}[htbp]
\caption{Vaccination status}
\subfloat[Survey wave January 2021]{\includegraphics[height=2.3in]{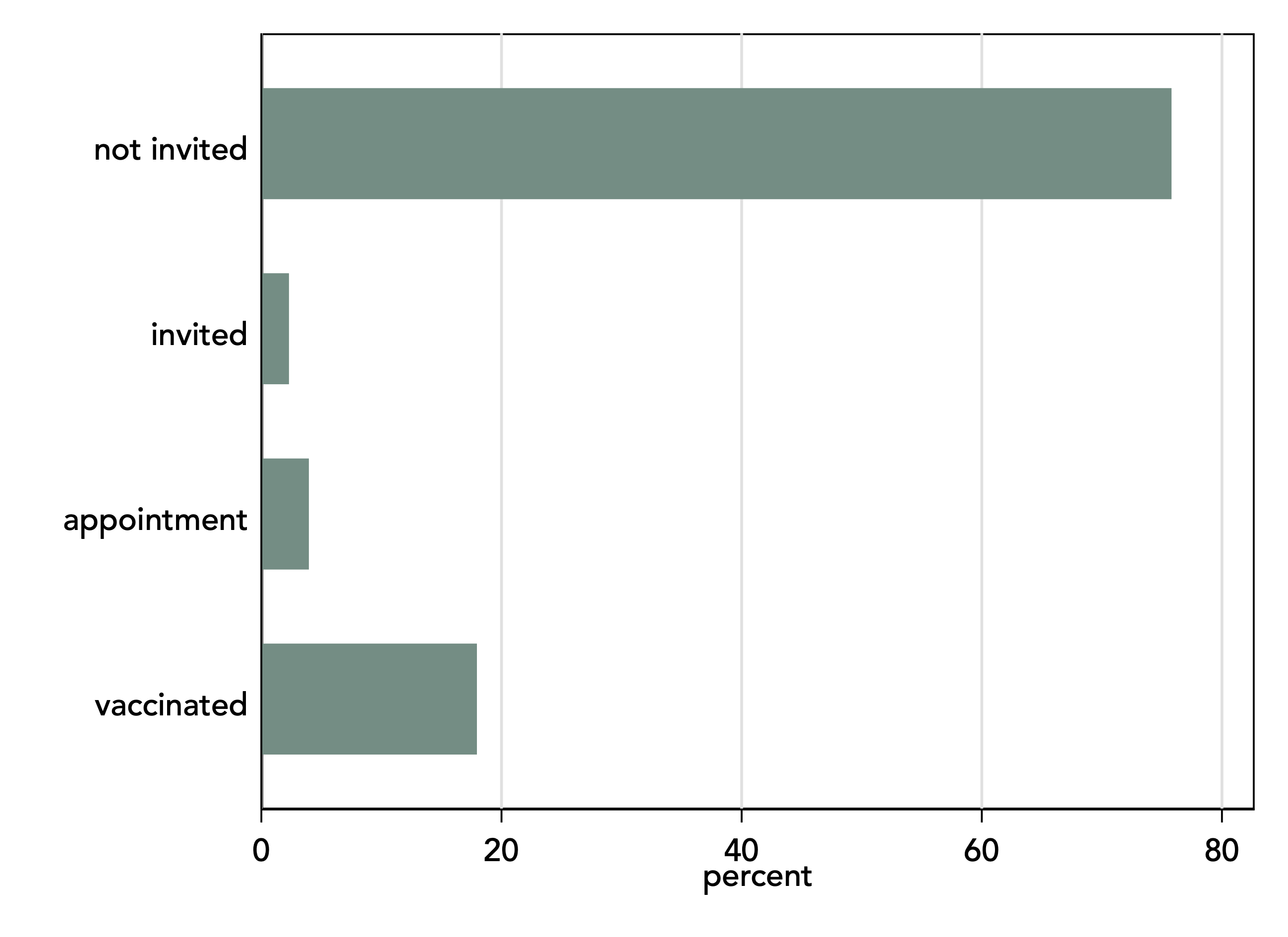}}
\subfloat[Survey wave March 2021]{\includegraphics[height=2.3in]{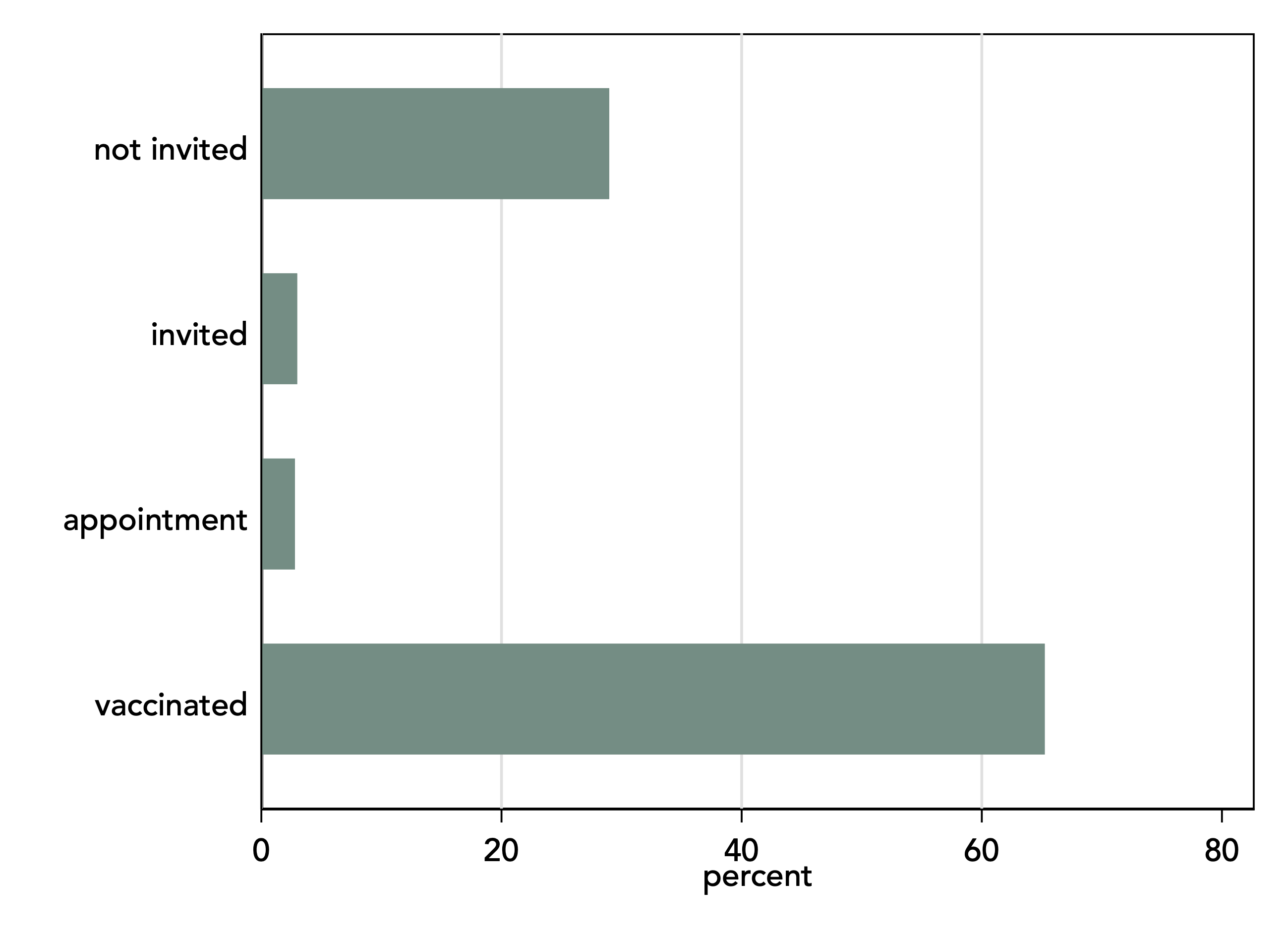}}
 \vskip 2mm 
\textit{Notes:} Population-weighted sample in the COVID-19 survey of \emph{Understanding Society}. The figure provides information on the percentage of (i) individuals not invited for vaccination, (ii) invited without an appointment or vaccination, (iii) invited with appointment and (iv) vaccinated, by survey wave. 
\label{fig:vax-status}
\end{figure}

\newpage
\begin{figure}[htbp]
\caption{Psychological well-being (GHQ-12) and COVID-19 incidence: 2019-2021.}
\begin{center}
\includegraphics[height=3.2in]{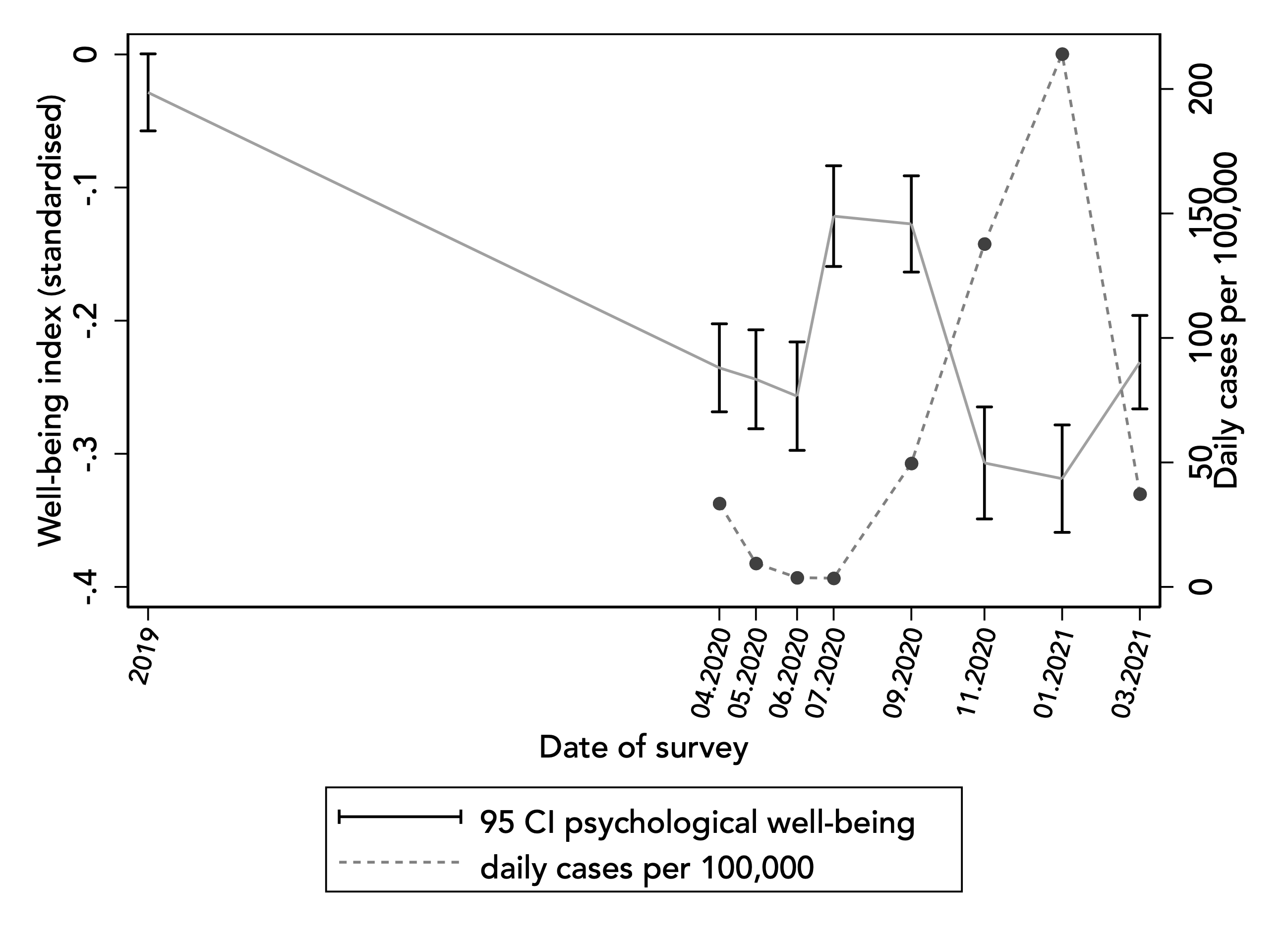}
 \end{center}
\caption*{\textit{Notes:} Population-weighted mean and 95\% confidence intervals at survey dates from the main survey and the COVID-19 survey of \emph{Understanding Society}. Information on daily cases from Public Health England.}
 \label{fig:timelines}
\end{figure}

\begin{figure}[htbp]
\caption{COVID-19 risk assessment and subsequent probability of testing positive}
\subfloat[Perceived COVID-19 risk]{\includegraphics[height=2.5in]{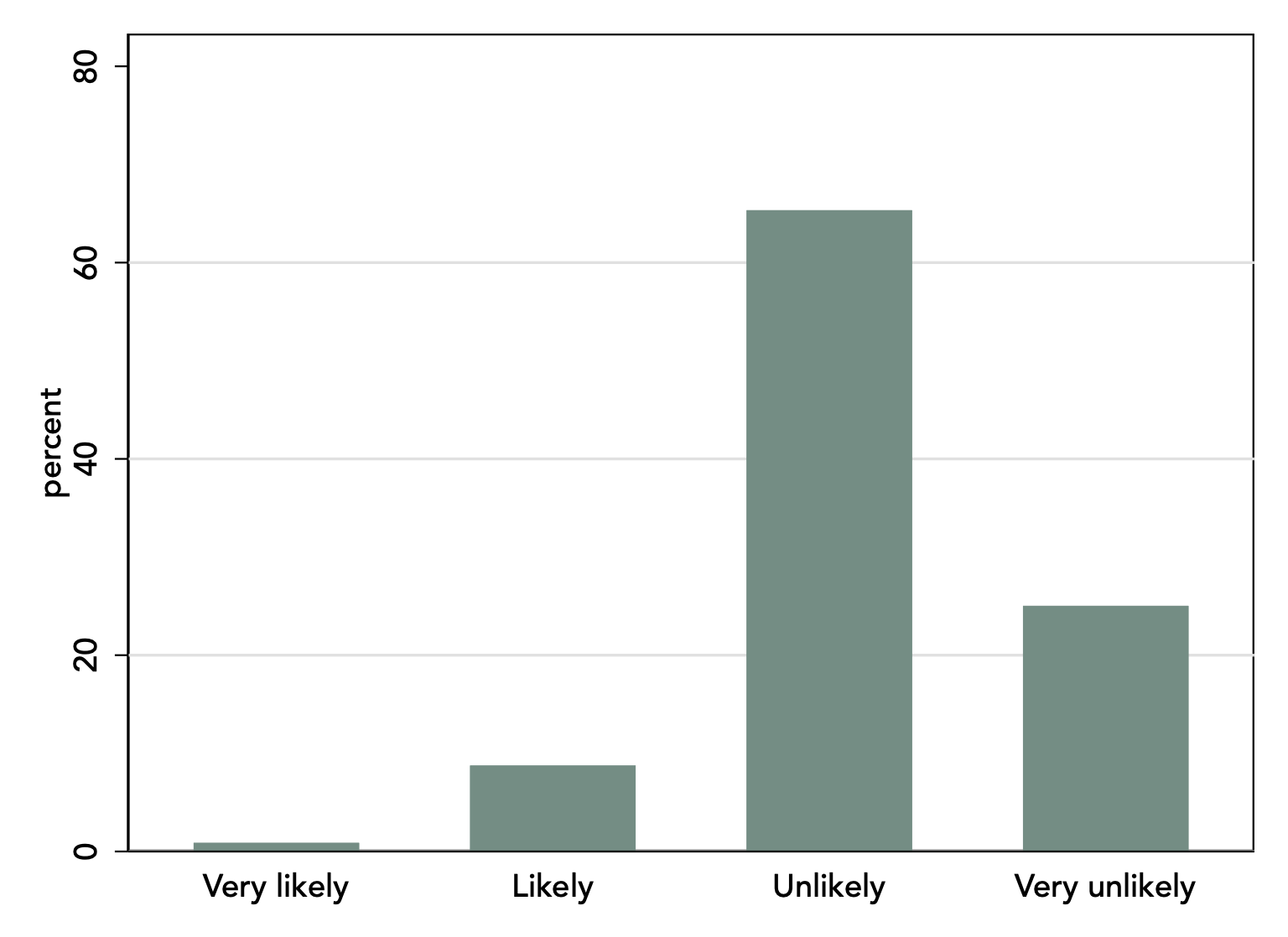}}
\subfloat[Testing positive by perceived COVID-19 risk ]{\includegraphics[height=2.5in]{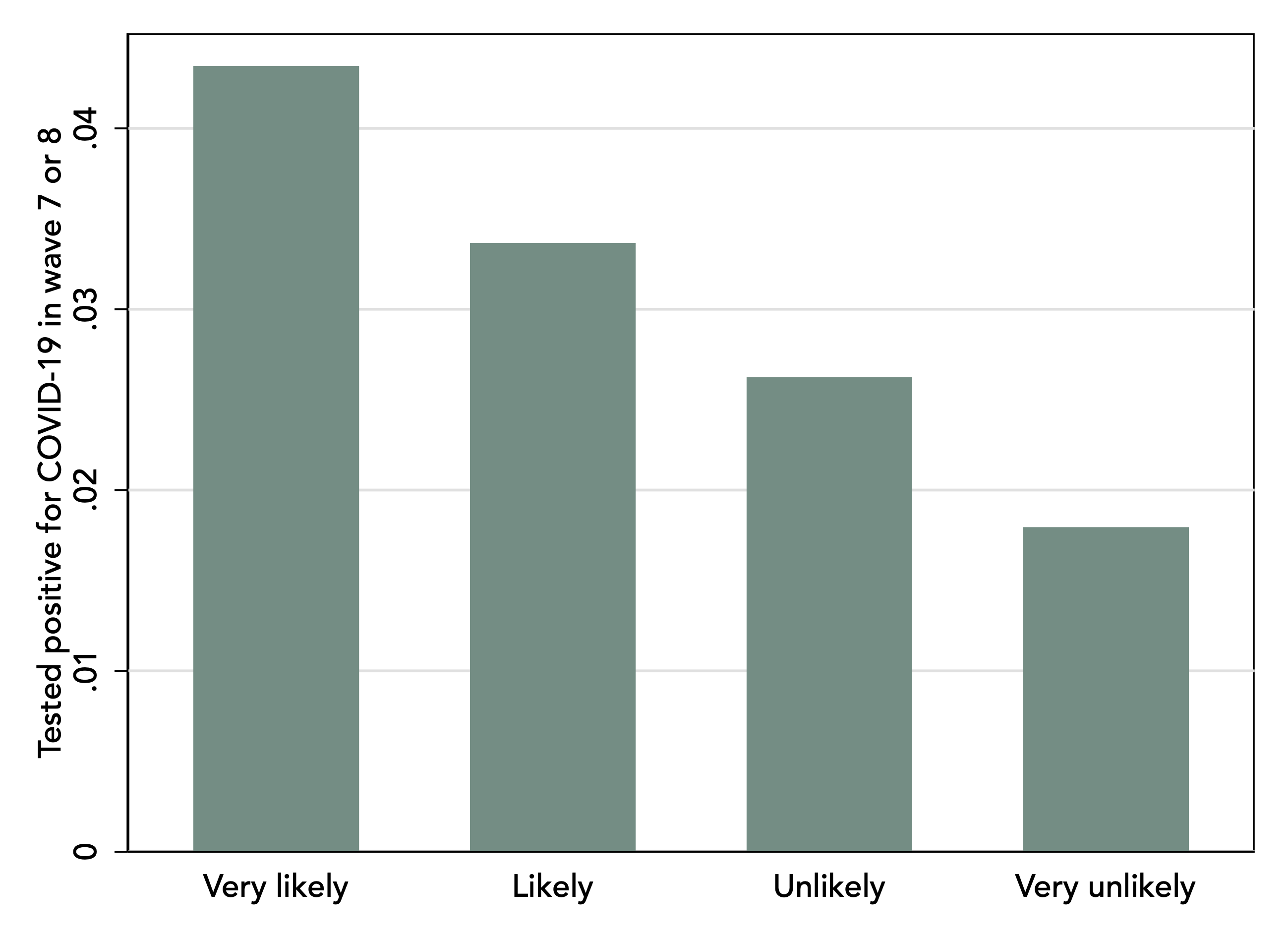}}
\\
\caption*{\textit{Notes:} Population-weighted sample in the COVID-19 survey of \emph{Understanding Society}. In panel (a), perceived COVID-19 risk is measured in wave 6 (November 2020) based on the following question: "In your view, how likely is it that you will contract COVID-19 in the next month?". In panel (b), we show the share of individuals who reported COVID-19 positive test by perceived COVID-19 risk in the previous wave.}
\label{fig:covid-risk}
\end{figure}

\begin{figure}[htbp]
 \caption{Reasons for taking or not taking the vaccine as in November 2020.}
\subfloat[What would be your main reason for taking \\ the vaccine?]{\includegraphics[height=2.3in]{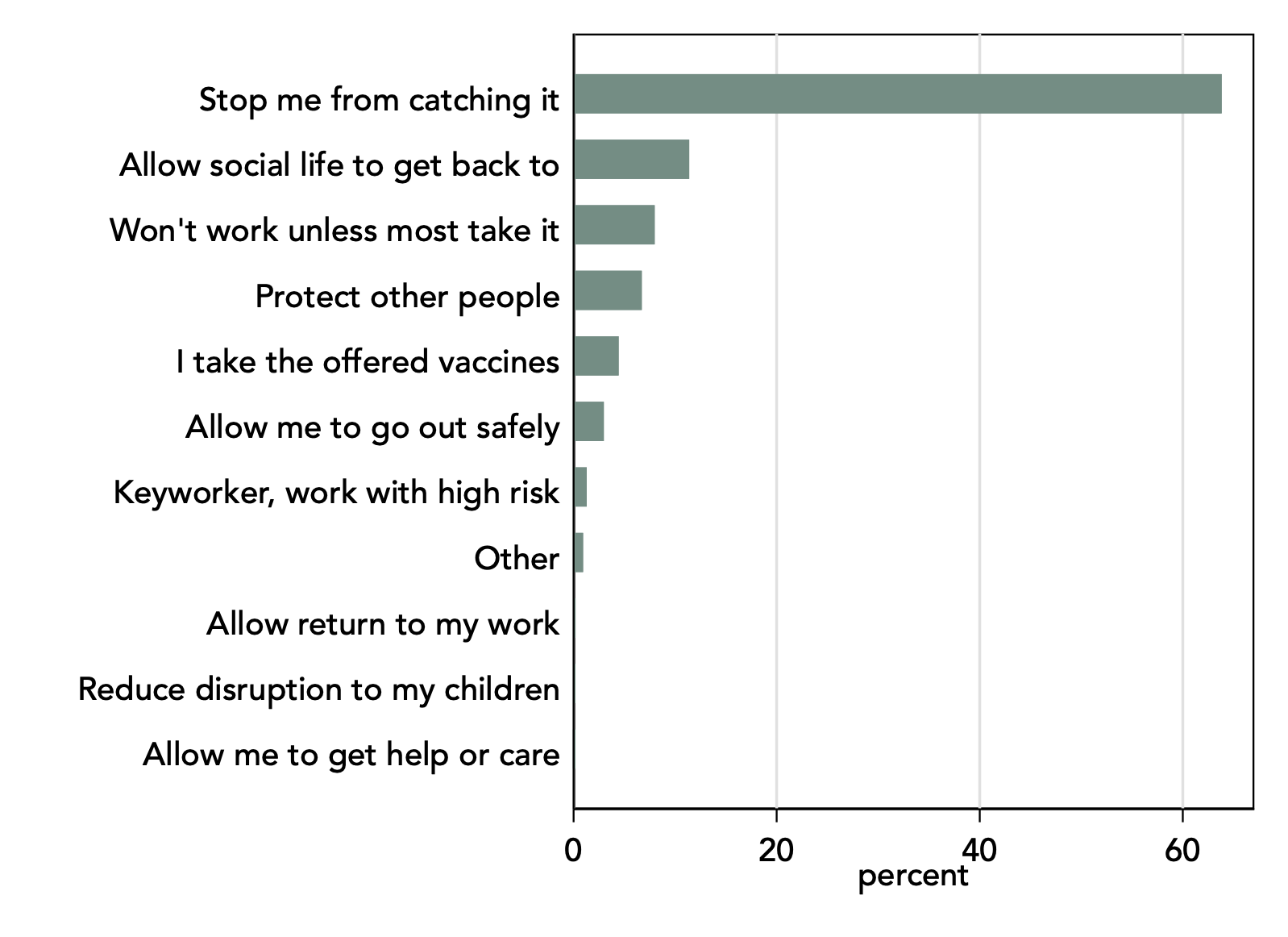}}
\subfloat[What is the main reason you would not take \\  the vaccine?]{\includegraphics[height=2.3in]{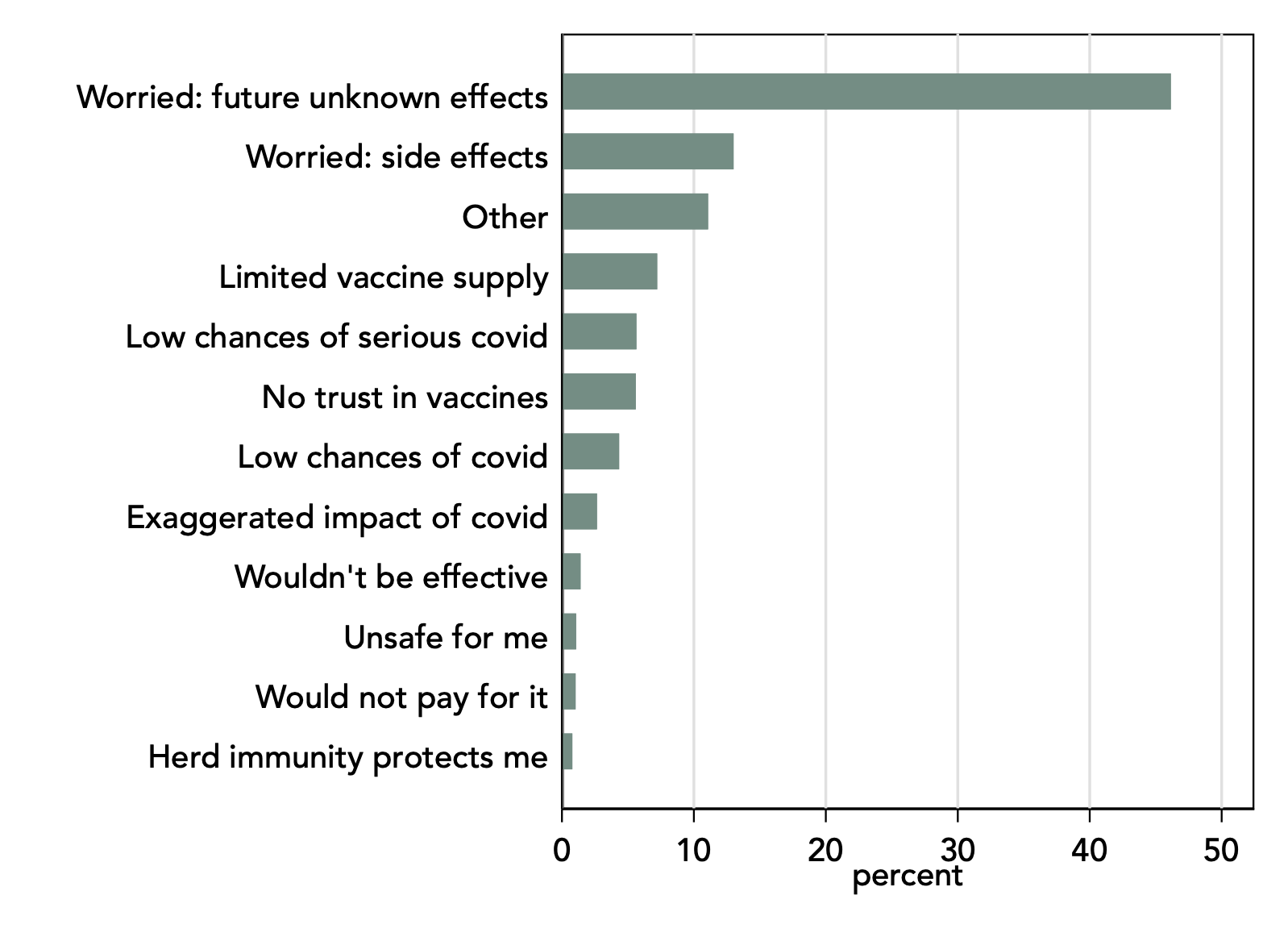}}\\
\textit{Notes:} Authors' calculation using data from \emph{Understanding Society} COVID-19 survey. The sample includes respondents aged 40-80, population-weighted in wave 6 (November, 2020).
\label{fig:vax-reasons}
\end{figure}

\begin{figure}[htbp]
\caption{COVID-19 vaccine attitudes}
\begin{center}
\subfloat[November 2020 (Wave 6), N = 11,853]{\includegraphics[height=2.3in]{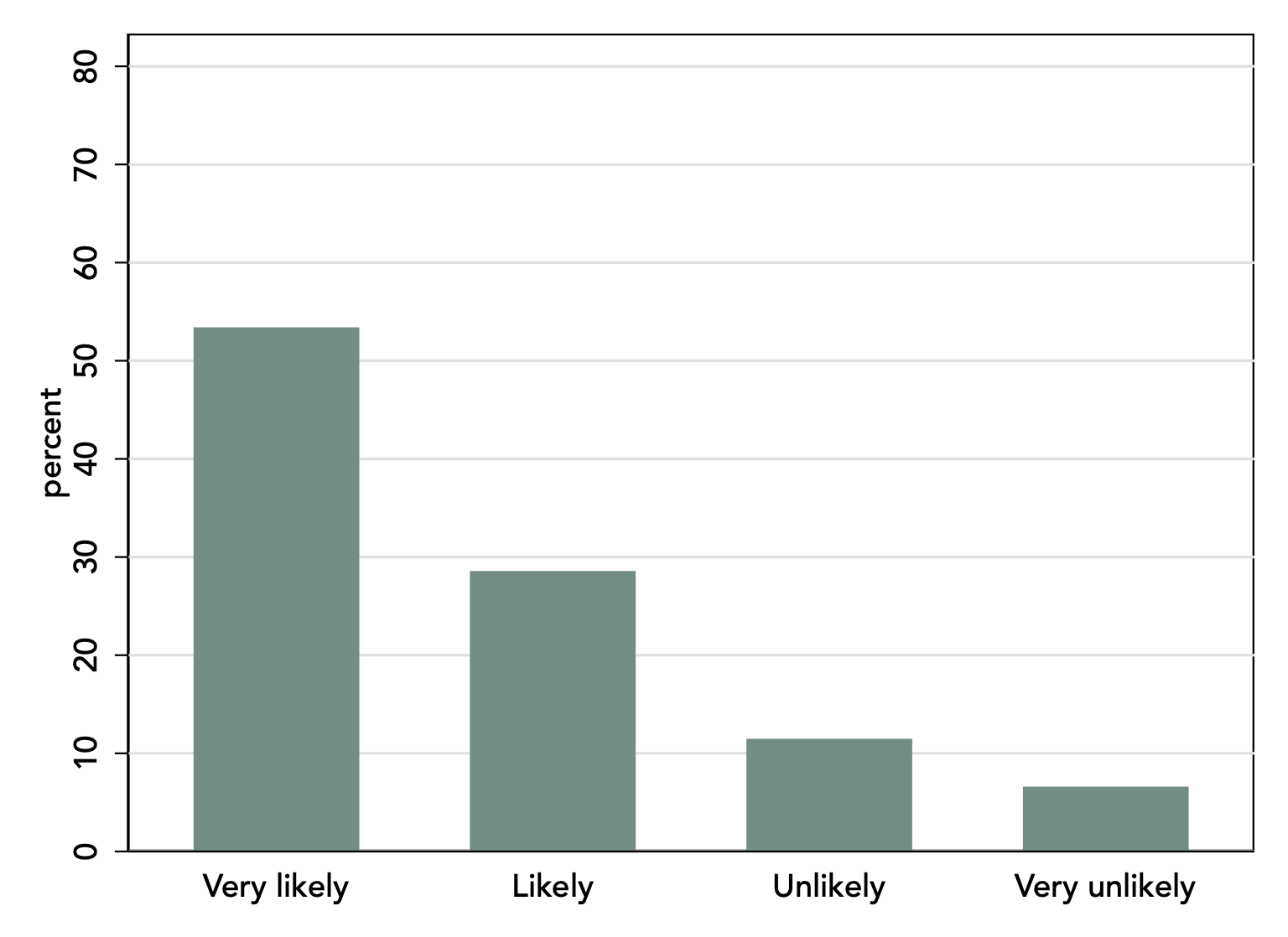}} \\
 \end{center}
 \vskip 2mm 
\caption*{\textit{Notes:} The table provides population-weighted information on the reply to the question "When you are offered the coronavirus vaccination, how likely or unlikely would you be to take it?". The sample includes all survey respondents in survey waves November 2020, January and March 2021.}
\label{fig:vax-attitudes}
\end{figure}

\begin{figure}[htbp]
\caption{Uptake heatmap of the vaccination roll-out by weeks and within age groups}
\begin{center}
\includegraphics[scale=.35]{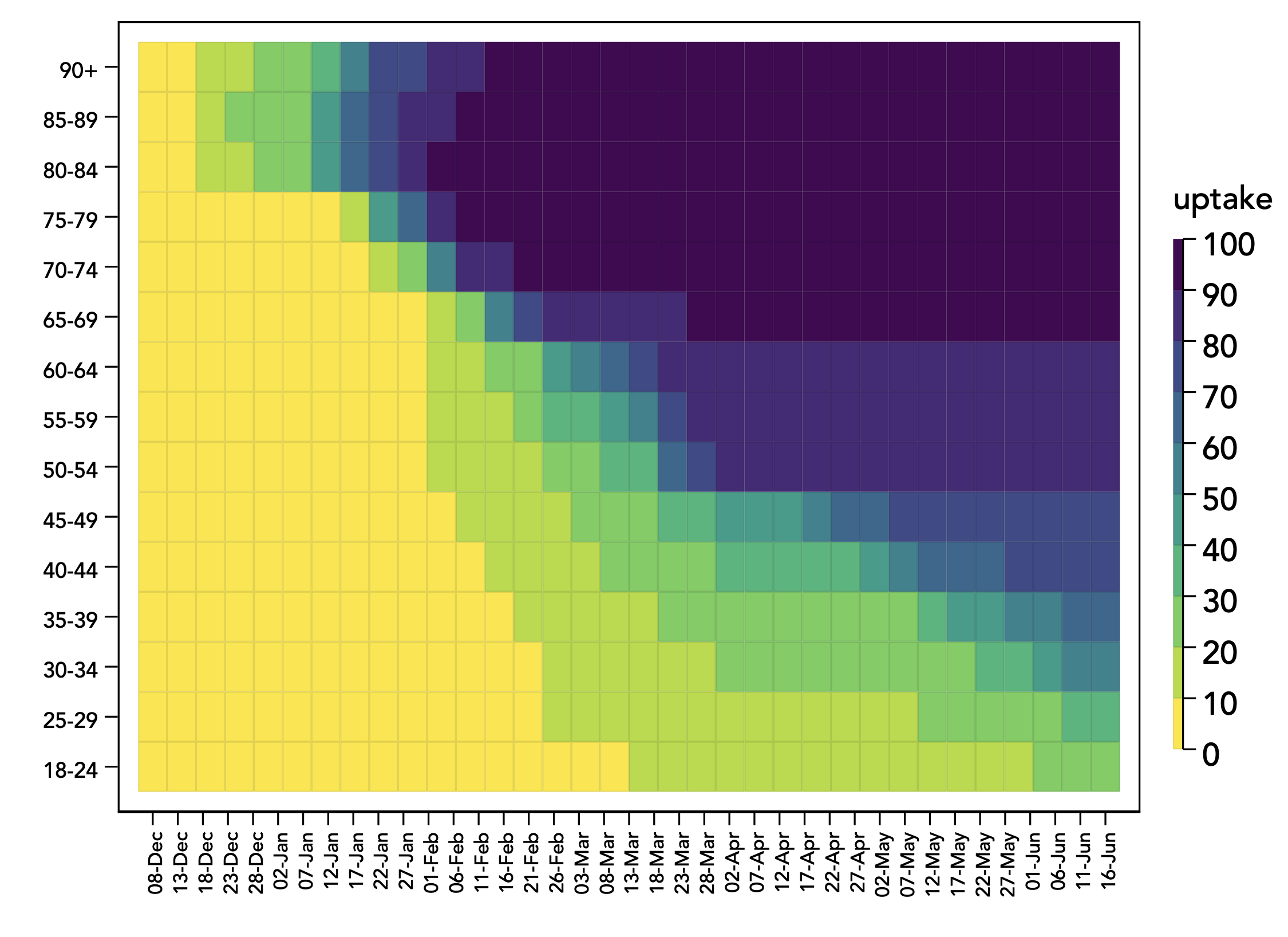}
 \end{center}
 \vskip 2mm 
\textit{Notes:} Total percentage of people who have received the 1st vaccination, by age group and vaccination date. Source: Public Health England, data from: https://coronavirus.data.gov.uk/
\label{fig:uptake-heatmap}
\end{figure}

\begin{figure}[htbp]
\caption{Uptake timeline by age groups.}
\subfloat[Cumulative date of vaccination, individuals aged 60-79.]{\includegraphics[height=2.4in]{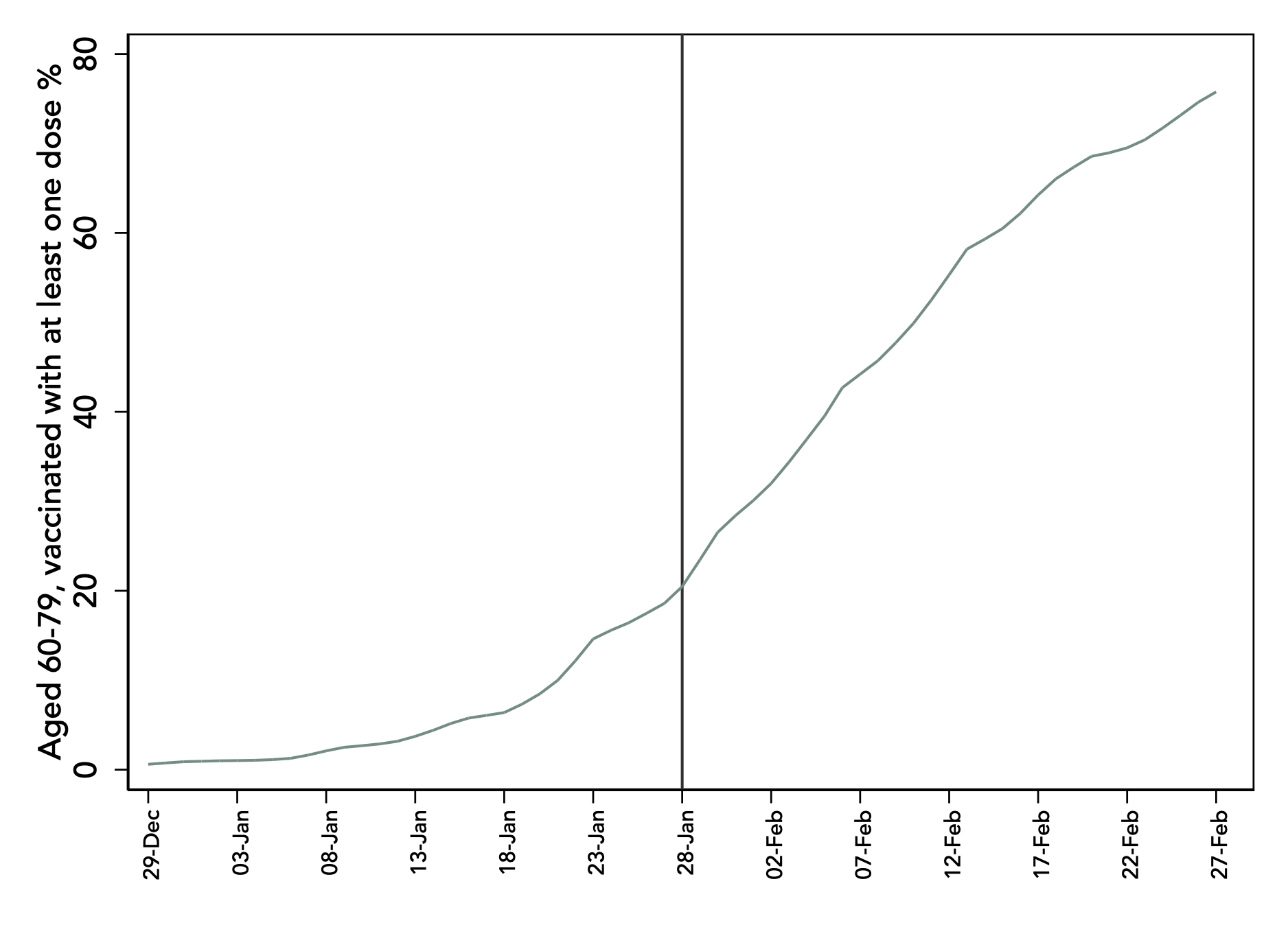}}
\subfloat[Cumulative date of vaccination, individuals aged 40-59.]{\includegraphics[height=2.4in]{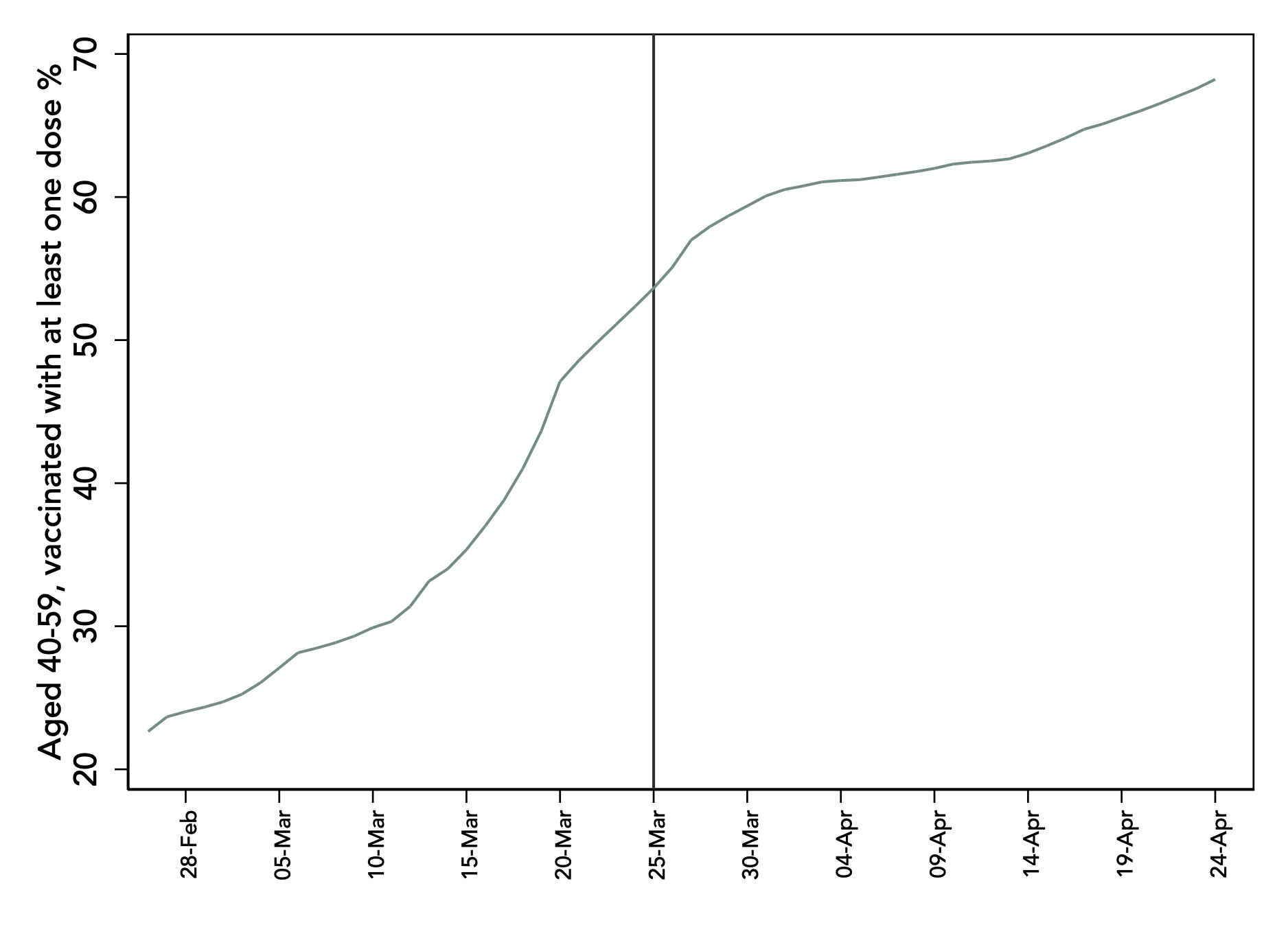}}
\\
\caption*{\textit{Notes:} Proportion of people who has been vaccinated with at least one dose of the vaccine, by date and age group. Source: Public Health England.}
 \label{fig:uptake_by_age}
\end{figure} 

\begin{figure}
\caption{Evolution of well-being for different age groups.}
\begin{center}
\includegraphics[height=2.5in]{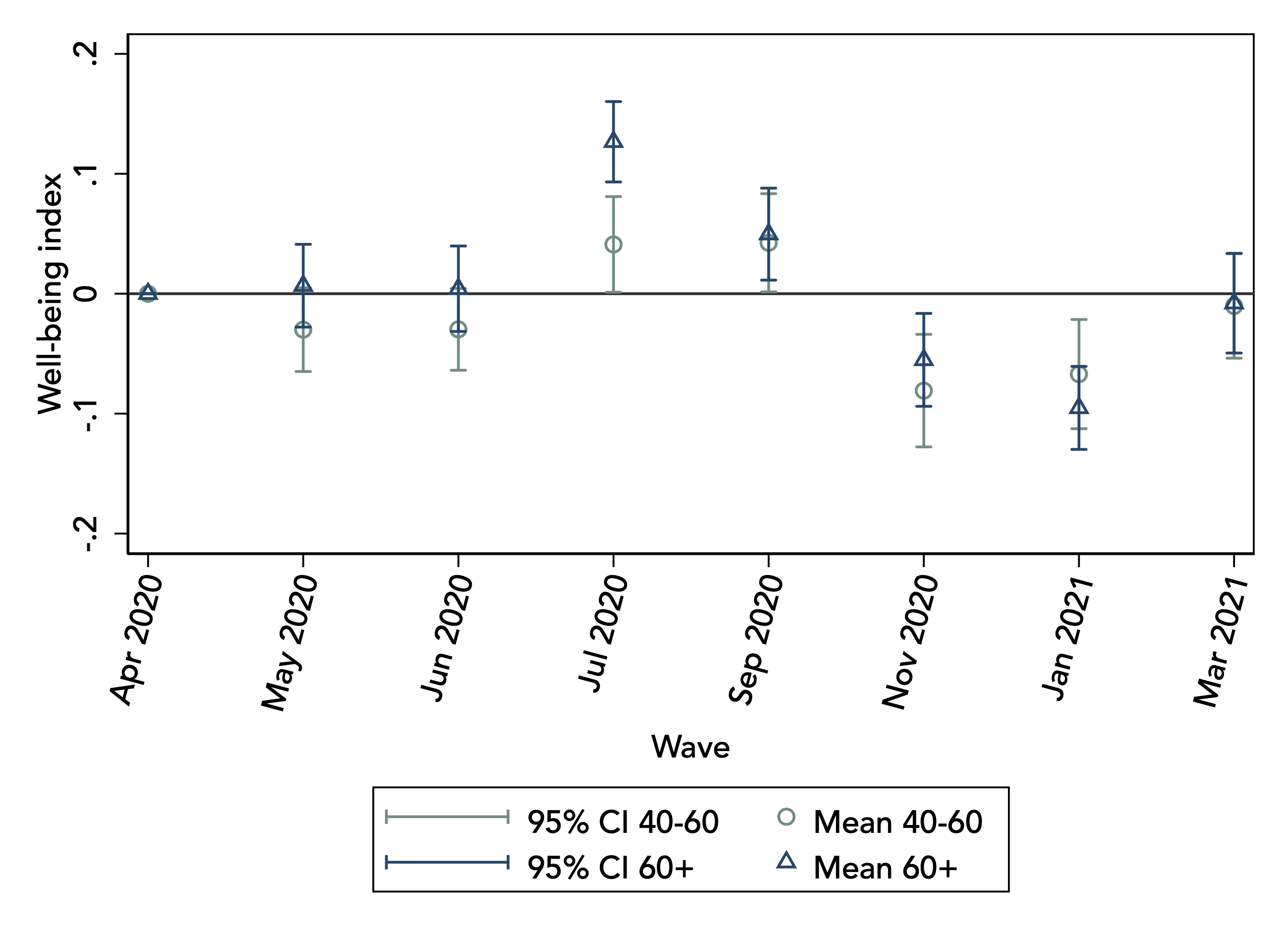} 
\end{center}
\textit{Notes:} Authors' calculation, using a OLS regression with individual fixed effects. The omitted group is the April 2020 survey wave and individuals under 40.
\label{fig:trend_by_age}
\end{figure}

\begin{figure}[htbp]
\caption{Event study - Impact of invitations on psychological well-being (GHQ-12)}
\begin{center}
\includegraphics[height=2.5in]{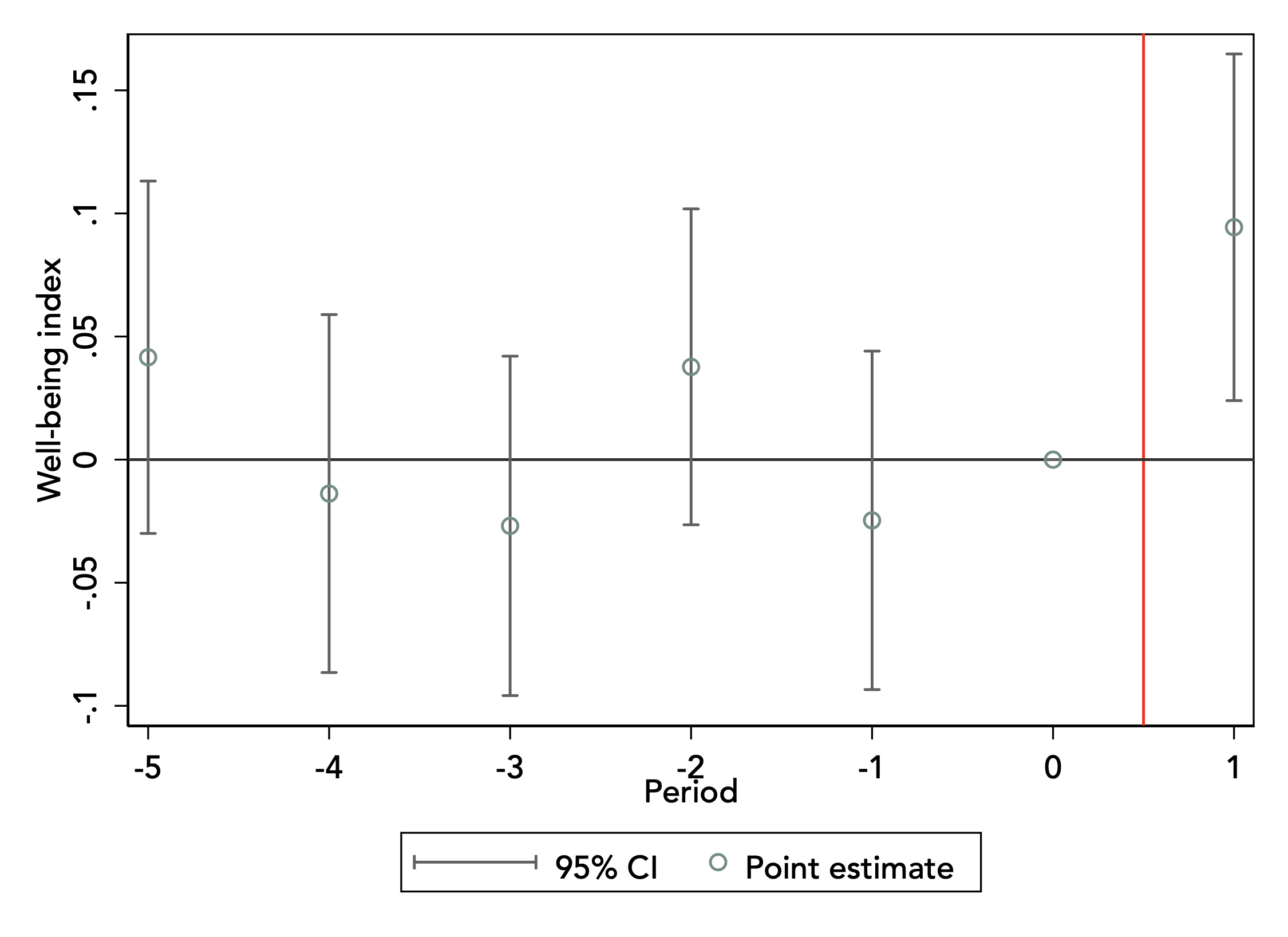} \\
\end{center}
\caption*{\textit{Notes:} The figure reports the estimates from the event study analysis of the IV-DID. The treatment group includes individuals aged 61-80 who had been invited for vaccination by the time of the January 2021 survey and individuals aged 40-60 who had been invited before the March 2021 survey. For individuals age 61-80 (40-60), we denote the November 2020 (January 2021) survey wave as the baseline period, and index all waves relative to that one. For more information see equation \ref{eq:eventstudy}).}
 \label{fig:event_study_invitation}
\end{figure}

\begin{figure}
\begin{center}
\caption{Impact of vaccination, by socio-economic group.}
\label{fig:coef_het}
\includegraphics[height=3.5in]{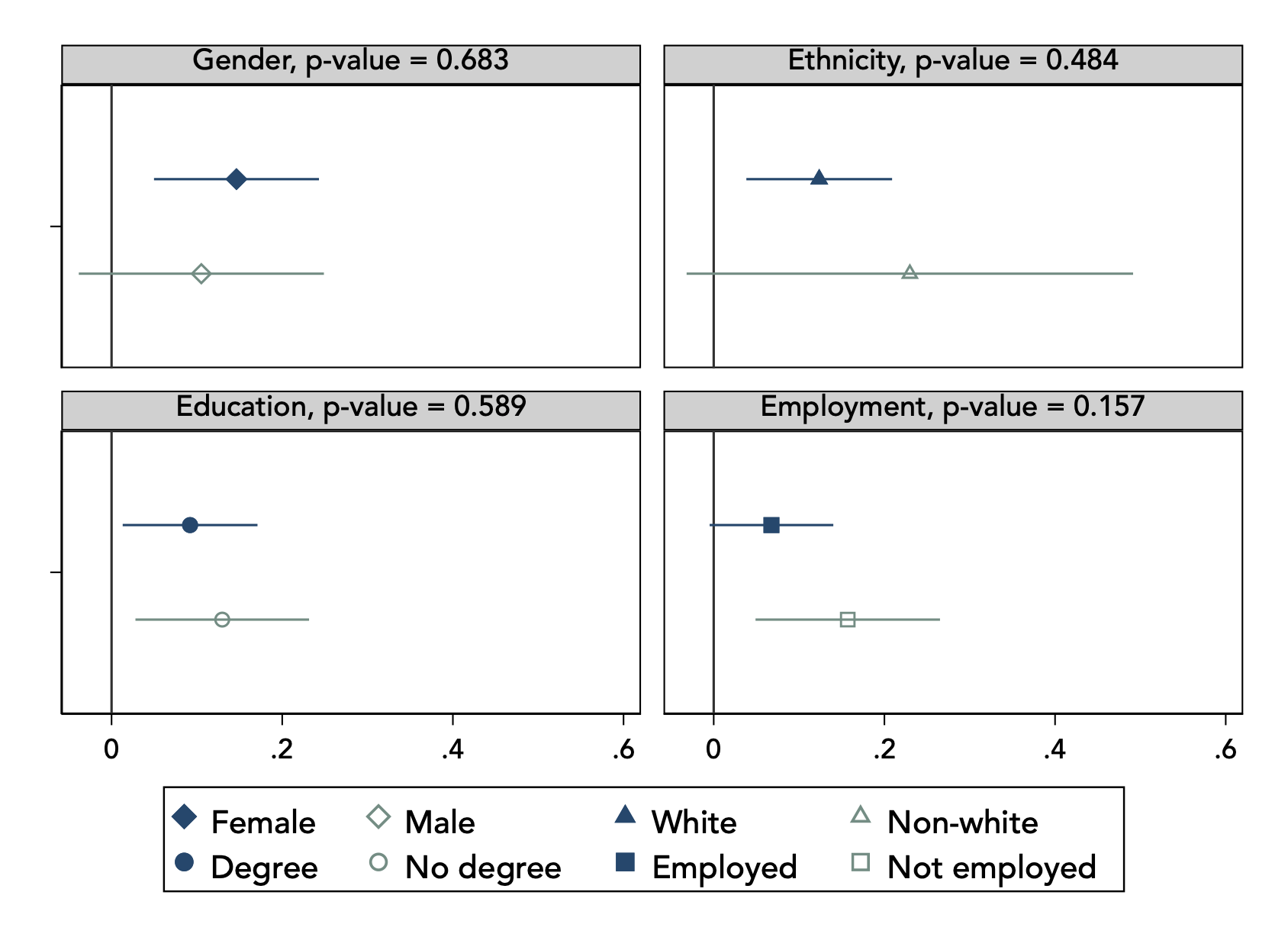} \\
\includegraphics[height=3.5in]{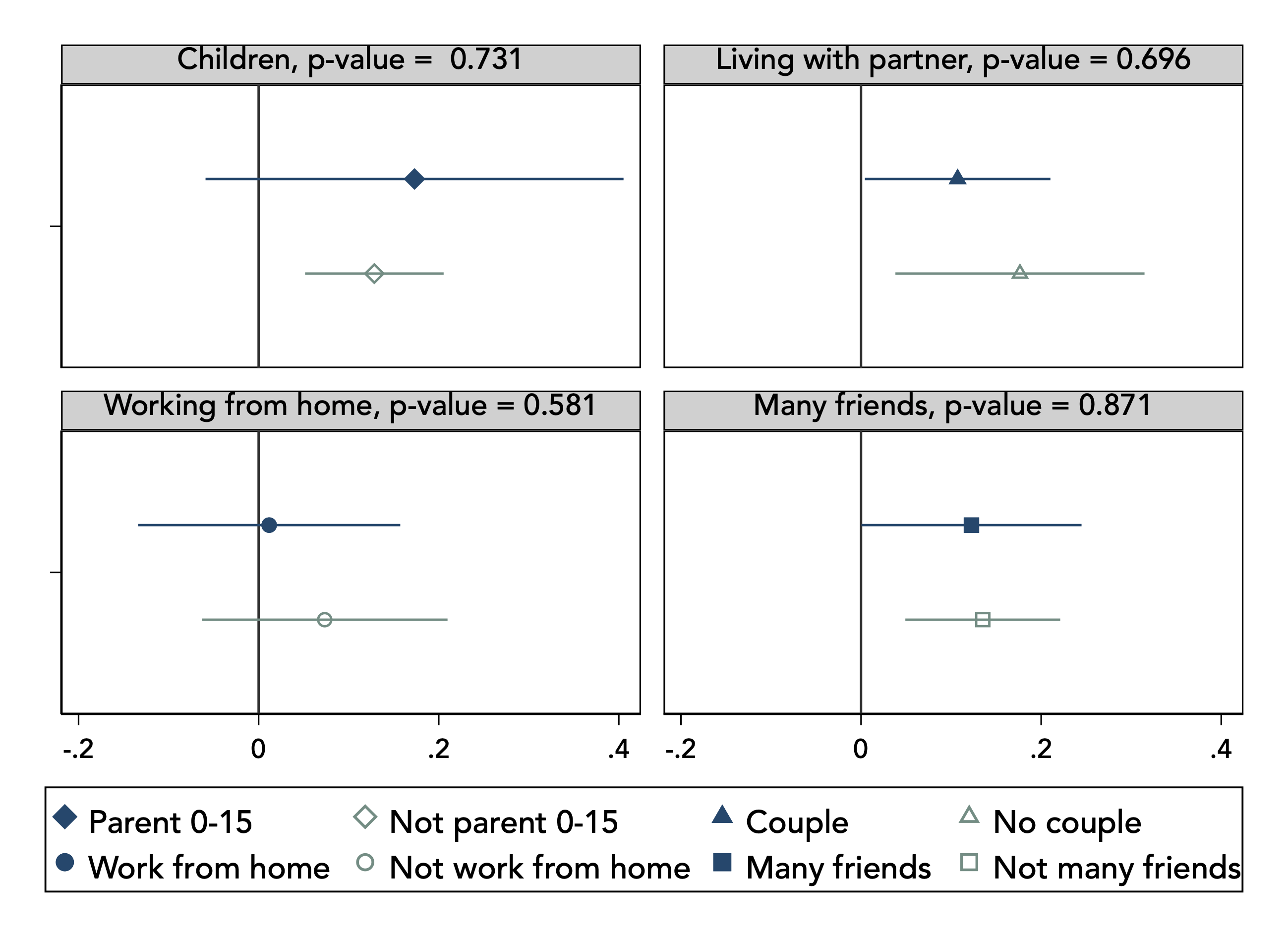}
\end{center}
\textit{Notes:} Impact of vaccination on psychological well-being (GHQ-12) for different socio-economic groups. The coefficient plots show the result from the 2SLS estimation of the vaccination on psychological well-being, in sub-samples according to the heterogeneity characteristic. The p-values are derived from a 2SLS estimation of the joint sample with an interaction of the endogenous regressor, the instrument and the fixed effects. Population-weighted sample in the COVID-19 survey of Understanding Society.
\end{figure}

\begin{figure}
\begin{center}
\caption{Histogram of age}
\label{fig:rdhistogram}
\includegraphics[height=3.5in]{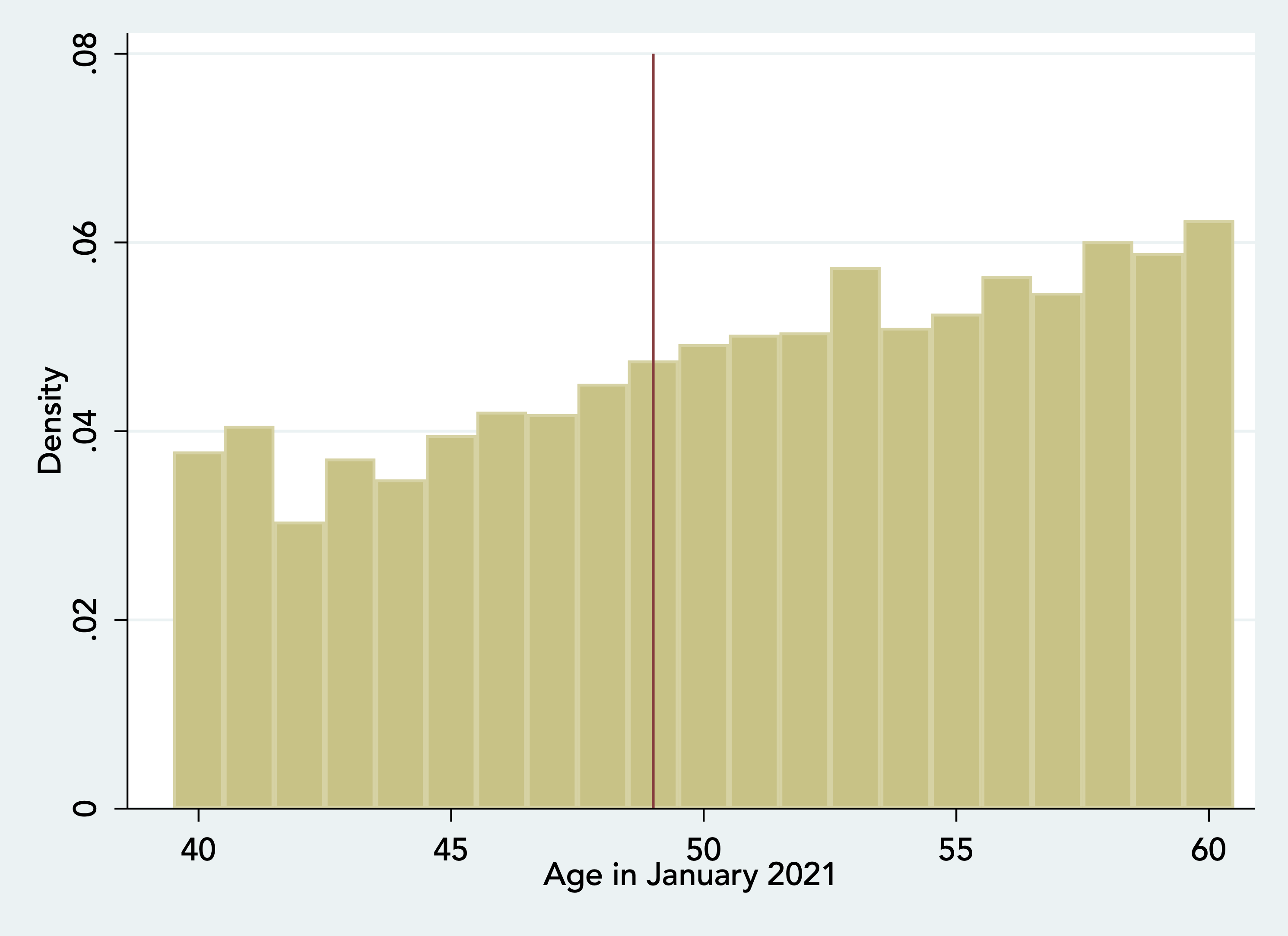} \\
\end{center}
\textit{Notes:} Histogram of age for individuals in the March 2021 survey wave, excluding Health and Social workers. The RD manipulation test proposed by \cite{cattaneo2020simple} rejects the existence of a discontinuity at the 49 years threshold (p-value=0.93).
\end{figure}

\end{appendices}

\end{document}